\documentclass[12pt]{article}
\pdfoutput=1
\usepackage{hyperref}
\usepackage{bm}
\usepackage{graphics}
\usepackage{graphicx}
\usepackage{setspace}
\usepackage[authoryear]{natbib}
\usepackage{amsmath,amsthm,amssymb}
\usepackage{titlesec} 
\usepackage{algorithm}
\usepackage{algorithmic}
\usepackage{color}
\onehalfspace
\DeclareMathOperator*{\argmax}{argmax} % thin space, limits underneath in displays
\DeclareMathOperator*{\argmin}{argmin} % thin space, limits underneath in displays

\title{\vspace{-15mm}\fontsize{24pt}{10pt}\selectfont\textbf{P-spline smoothing for spatial data collected worldwide}}

\author{
\large
\textsc{Fedele Greco},  
\textsc{Massimo Ventrucci}\thanks{Corresponding author. Email: \texttt{massimo.ventrucci@unibo.it}}   ,
 \textsc{Elisa Castelli}
\\[2mm] % Your name
\normalsize Department of Statistical Sciences, University of
Bologna, Bologna, Italy\\
\normalsize Institute of Atmospheric Sciences and Climate (CNR-ISAC), Italy
\vspace{-5mm}
}
\date{}

\begin{document}
\maketitle 

\begin{abstract}
%\textbf{Keywords:}
Spatial data collected worldwide at a huge number of locations are frequently used in environmental and climate studies. Spatial modelling for this type of data presents both methodological and computational challenges. In this work we illustrate a computationally efficient non parametric framework to model and estimate the spatial field while accounting for geodesic distances between locations. The spatial field is modelled via penalized splines (P-splines) using intrinsic Gaussian Markov Random Field (GMRF) priors for the spline coefficients. The key idea is to use the sphere as a surrogate for the Globe, then build the basis of B-spline functions on a geodesic grid system. The basis matrix is sparse and so is the precision matrix of the GMRF prior, thus computational efficiency is gained by construction. We illustrate the approach on a real climate study, where the goal is to identify the Intertropical Convergence Zone using high-resolution remote sensing data.
%\textbf{Keywords:}
\end{abstract}

\section{Introduction}
\label{sec:intro}
High-resolution spatial data collected worldwide, usually by means of remote sensing techniques, are widely spread in environmental and climate studies: most of the statistical methods developed in modelling this kind of data use the sphere as a surrogate for the Globe. Modelling data collected at a global scale presents both methodological and computational challenges. The traditional toolkit of the spatial data modeller facing geostatistical datasets and aiming to make prediction at unmonitored locations suggests to apply kriging techniques (see, e.g., \cite{banerjee-2015}). These rely on the assumption of a smooth Gaussian Random Field (GRF), continuous in space but only observed at a discrete set of points, any finite realization of it being generated by a multivariate Gaussian distribution. The covariance structure of this distribution is specified via a spatial covariance function. The practice is largely dominated by spatial covariances defined on Euclidean distances, such as the Mat\'{e}rn family, thus a preliminary step in the analysis is the projection of the 3d Cartesian coordinates (on the Earth surface) over a 2d coordinate space. Standard choice is to work with geographic coordinates (latitude-longitude), but other mappings can be used. \cite{banerjee-geo} provides a review of such mappings and discusses the impact of the chosen metric on spatial prediction via kriging. The traditional toolkit outlined above  presents two main difficulties when modelling high-resolution data observed over a spherical domain.

The first issue is that the process of spatial prediction needs to be coherent with the geometry of the sphere. Using a planar metric over a 2d projection is inappropriate because it generates spurious anisotropy and non-stationarity of the covariance function (\cite{banerjee-geo}). 
The geodesic (aka great circle) distance, i.e. the length of the shortest path between two points over the surface of a sphere, is a natural candidate for measuring distances over a spherical domain. 
However, using great circle distances in a Mat\'{e}rn family does not necessarily guarantee a positive definite covariance \citep{gneiting-2013}. %\cite{huang-2011, jeong1-2015}. 
\cite{banerjee-geo} studied via simulation the behaviour of different metrics regarding estimation of the Mat\'{e}rn covariance parameters on a region as large as Colorado, finding a substantial impact of the chosen metric on the range of the correlation function. This means that with data collected on larger regions on Earth (e.g. the whole Globe), biased estimation of the underlying field has to be expected to some extent, when covariance functions built on Euclidean distances are used. A large number of papers have tackled this issue by essentially proposing new models for data on a spherical domain, both in a parametric and non-parametric framework. 

In the parametric setting, several papers focused on building valid stochastic processes for the sphere, see, e.g., \cite{jun-stein-2007, jeong2-2015, heaton-2014} and references therein. The stochastic partial differential equation (SPDE) approach has grown a lot of attention recently \citep{lindgren-2011, sangalli-2013,sigrist-2015}. This approach builds a GRF as the finite element solution of a particular SPDE, an idea that can be generalized to different types of manifolds including the sphere. The paper by \cite{lindgren-2011} focuses in particular on the computational properties of the SPDE approach, deriving an approximated solution in terms of a Gaussian Markov Random Field (GMRF), instead of a GRF, in order to gain computational speed. 

In the non-parametric setting, \cite{wahba-1981} firstly introduced smoothing splines on the sphere motivated by the analysis of weather data collected at a large number of stations around the world. Outside the spline realm, \cite{dimarzio-2014} presented local linear regression for spherical data, including smoothing of a scalar response on a spherical predictor as a particular case. \cite{wood-2017} discusses in detail the connection between spline smoothing and thin plate splines for the sphere, pointing out that low rank smoothers are also applicable to spherical data. Although low rank smoothers allows reduction of the number of parameters to estimate, the main role in alleviating the computational burden is played by the sparsity of the smoothing matrix, obtained by using local basis functions, i.e. non null over a limited domain. B-splines are local functions built upon joint polynomials connected at knots and are applied in different contexts, e.g. in penalized spline (P-spline) non-parametric regression \citep{eilers-1996} or in the SPDE approach mentioned above. %In the spherical context B-splines have been used for data interpolation in \cite{lai-2009}. 
In this work, the computational properties of B-splines are exploited.  

Indeed, the second difficulty concerning the application of kriging techniques to high-resolution global datasets is purely computational. Continuous covariance functions used in geostatistics imply a dense covariance structure for the underlying GRF. When the number of data locations $n$ is large, this modelling framework becomes impractical because of the need to invert large dense matrices, with a computational cost increasing in cubic order with $n$. Statistics literature on the \emph{big n problem} has grown incredibly fast in the last decade, mostly motivated by the increasing availability of high resolution remote sensing data for environmental studies. Some of the models for large data that can be implemented in a fully Bayesian hierarchical setting (for a review see \cite{banerjee-2017}) are based on a low-rank representation of the field \citep{wikle-1999, banerjee-2008}. Other proposals seek a sparse representation of the covariance, like tapering \citep{furrer-2006}, or of the precision, like in the SPDE approach (i.e. relying on the Markov properties of the graph underlying the model \citep{rue-2005}). The fully Bayesian framework presented in this paper follows both directions, as it is built on a low-rank representation and exploits the sparsity induced by a GMRF prior.

We propose a computationally efficient non-parametric approach to estimate the spatial field underling data on the sphere that properly accounts for geodesic distances between locations. Our method is based on a low-rank P-spline smoother to gain flexibility w.r.t. parametric models.
% such as the Mat\'{e}rn models derived via SPDE or the covariance models designed for the sphere. 
The main contribution of this work is the extension of the P-spline model for smoothing data collected over a spherical domain. The model is built on a set of bivariate B-splines computed on a Geodesic Discrete Global Grid (GDGG) system \citep{sahr-2003}, yielding a quasi-regular triangular mesh over the Globe. Geodesic grids have been used in spatial statistics to create flexible multi-resolution models implemented in a likelihood based inferential framework \citep{cressie-2008,nychka-2012}. In contrast to the latter works, in this paper we follow a fully Bayesian approach and fit the model using an efficient Gibbs sampler, exploiting sparsity of the basis matrix and of the precision of the GMRF prior. We illustrate the method on a real climate study, where the goal is to identify the \emph{Intertropical Convergence Zone} (ITCZ) from high-resolution remote sensing data collected worldwide over sea, with missing data occurring over land.

The rest of the paper is organized as follows. In section \ref{sec:motivating} the dataset and application goals are described. Section \ref{sec:model} presents our proposal for smoothing data over the sphere that we dub \emph{Geodesic P-splines}. Section \ref{sec:results} illustrates the method on a climate case study, focusing on the detection of the ITCZ. A discussion is provided in Section \ref{sec:discussion}.

\section{Motivating example}
\label{sec:motivating}

Our interest in geodesic P-splines is motivated by a climate case study aiming at investigating the location of the ITCZ using satellite data. The ITCZ is a region of the atmosphere broadly located within the tropical belt where the north-east and
south-east trade winds converge, which is characterised by high cloudiness and severe convective precipitation \citep{Holton20121}. An important aspect regards seasonal variability in the ITCZ position: ITCZ is roughly located North of the equator in the boreal spring and summer, while it migrates to southern regions in autumn and winter. The location of the ITCZ affects duration and intensity of the wet and dry seasons at the tropics and plays a key role in the general circulation of the atmosphere: assessing its variability is crucial for improving global climate models. Moreover, understanding the long-term trend characterizing this phenomenon is crucial for monitoring changes in climate pattern on a global scale. 
%[ECCO COME ERA QUESTO PEZZETTO]The ITCZ position affects duration and intensity of the wet and dry seasons at the tropics and plays a key role in the general circulation of the atmosphere: assessing its variability is crucial for improving global climate models. The variability of the ITCZ is not solely related to climate features in the tropic belt, in fact understanding the long-term trend characterizing this phenomena is crucial for monitoring changes in climate pattern on a global scale. Another important aspect regards seasonal variability, as the ITCZ is roughly located North of the equator in the boreal spring and summer, while it migrates to southern regions in autumn and winter.

The phenomenon regulating the ITCZ behaviour cannot be measured directly, hence several studies have investigated it using some suitable proxy variable, like maximum precipitation \citep{zhang2001double}, wind field \citep{vzagar2011climatology}, vorticity and reflectivity of the clouds \citep{waliser1993satellite}. As a general feature, all these studies benefit from the increasing availability of satellite measurements. In this paper we focus on data from the infrared channels of the Along Track Scanning Radiometer (ATSR) instrument series, 
%(ALTS project https://earth.esa.int/web/sppa/activities/multi-sensors-timeseries/alts/about), 
that were in orbit from 1991 to 2012 for accurate retrieval of sea surface temperature. Recently, in the frame of the European Space Agency ATSR Long Term Stability project
%\citep{ALTS-project},
(\href{https://earth.esa.int/web/sppa/activities/multi-sensors-timeseries/alts/about}{https://earth.esa.int/web/sppa/activities/multi-sensors-timeseries/alts/about}), 
\cite{casadio2016total} developed the Advanced Infra-Red Water Vapour Estimator algorithm (AIRWAVE) for the retrieval of the Total Column of Water Vapour (TCWV) from the ATSR measurements. In this work we use TCWV as a proxy variable for locating ITCZ. 

\begin{figure}
\centerline{
\includegraphics[width=0.475\textwidth]{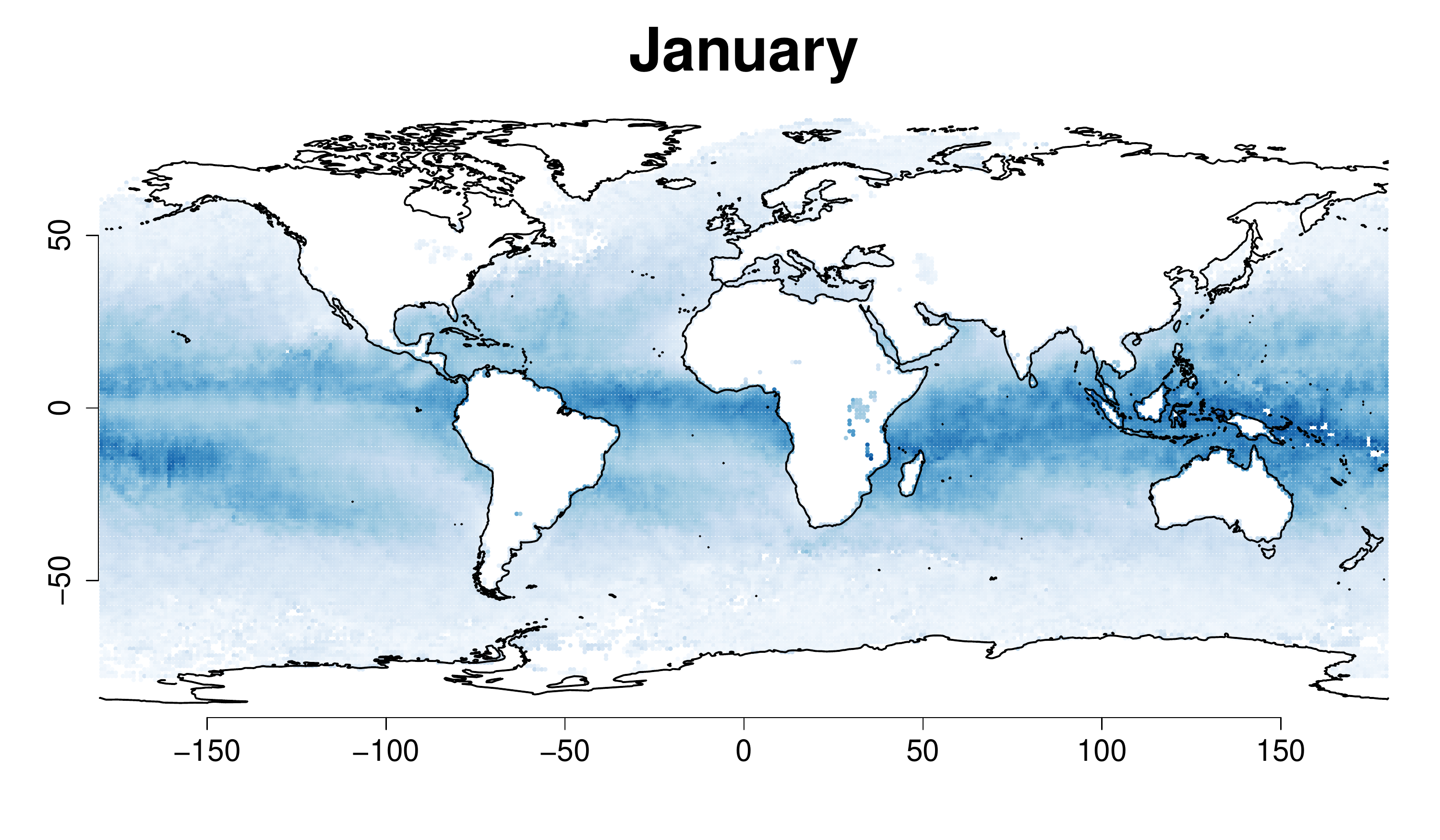}
\includegraphics[width=0.475\textwidth]{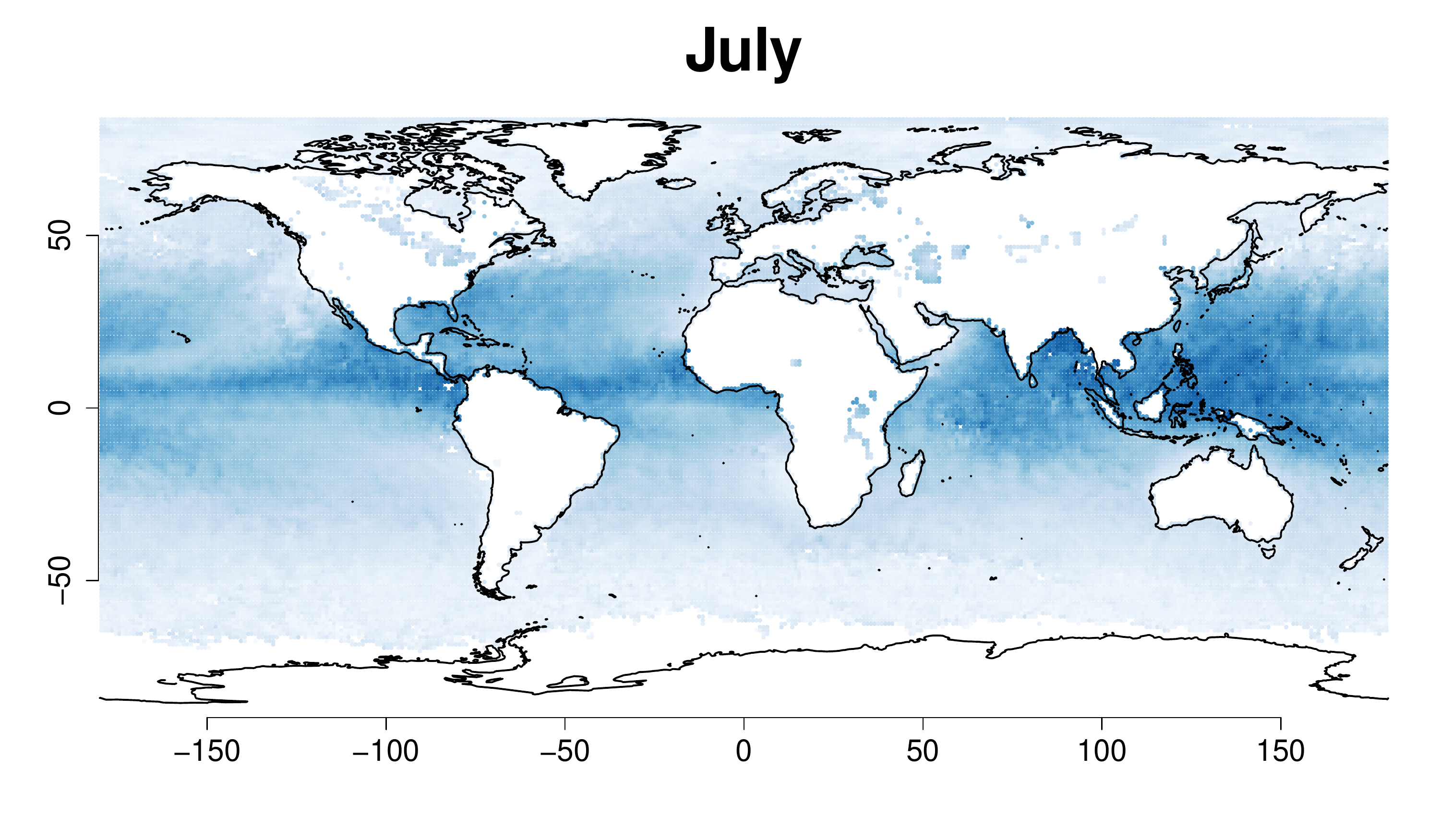}
\includegraphics[width=.05\textwidth]{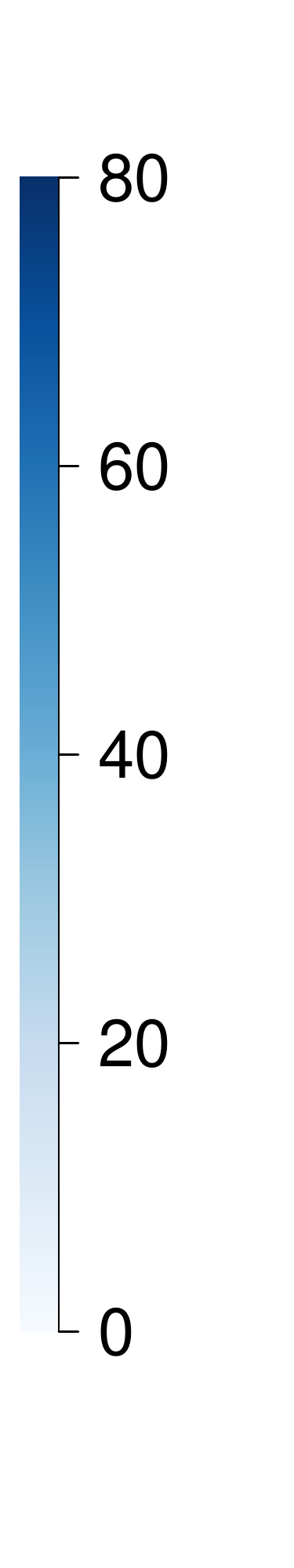}
}
\caption{TCWV data for January and July (unit measure, $Kg/m^2$). In general, TCWV measurements are available only over sea, as the data cannot be accurately retrieved over land; however, note that observations are still present in correspondence of wide lakes, e.g. the Great Lakes of North America and the Victoria lake.}
\label{fig:data}
\end{figure}

Data on TCWV regarding year 2008 were provided by the National Research Council - Institute of Atmospheric Sciences and Climate (CNR-ISAC), Italy. Data come as monthly averages of TCWV on a raster grid of $720$ columns (longitude values) and $360$ rows (latitude values), thus each cell covers half degree over latitude and longitude. In Figure \ref{fig:data}, data for January and July are displayed. The AIRWAVE algorithm provides reliable data over the sea and in clear sky conditions, thus TCWV observations are missing over land (broadly a third of the total number of cells), except in areas covered by lakes. 

The application goal is to estimate the ITCZ position and its uncertainty. We consider the TCWV data on the fine raster grid as point-level data observed at the centroid of each cell and focus on modelling the latent field of TCWV separately for each month, leaving spatio-temporal modelling to future work. The statistical challenges we tackle in this paper are related to efficient smoothing of large data to remove measurement error and to fast prediction at unmonitored locations. We believe that extension of Bayesian P-Splines to a spherical domain can be a valuable strategy because of its efficiency and computational stability.% and fastness. 
Bayesian inference provides immediate tools for ITCZ location, by analysing the joint posterior distribution of the latent field. In Section \ref{sec:results}, the ITCZ detection problem is addressed by searching for the latitudes where the TCWV latent field shows highest values. We provide a graphical output, by plotting the posterior probability that a point on the Earth belongs to the ITCZ.

%%%% FIGURE data: gennaio e luglio

\section{Geodesic P-splines}
\label{sec:model}

\subsection{Background on P-splines for spatial data}
\label{sec:classic-P-splines}

In the one dimensional setting, P-splines \citep{eilers-1996} are usually adopted to model the smooth effect of a covariate on the response as a linear combination of B-splines scaled by spline coefficients. Key features of this method are (a) equally-spaced univariate B-splines of certain degree $d$, these being non zero over a limited interval of the covariate domain, and (b) a penalty on the $r^{th}$ order differences between adjacent spline coefficients to control smoothness. The popularity of P-splines is due to numerical stability and flexibility in the modelling choices; e.g., the penalty order and the degree of the B-splines can be decided according to the application at hand. Higher-dimensional smoothers, suitable for modelling spatial data, can be constructed as tensor product P-splines \citep{eilers-2006}. In a frequentist framework, estimation is obtained via penalized maximum likelihood or iterative re-weighted least squares, with the smoothing parameter selected via cross validation or optimized over some information criterion. This method has become increasingly popular and is currently implemented in \texttt{R} packages such as \texttt{mgcv} \citep{wood-2017}. %We find a Bayesian approach more customary for the purpose of our case-study. 

In order to build the ground for our proposal we next revise spatial P-splines for data observed on a two-dimensional latitude-longitude plane following \cite{eilers-2006}. Let us assume $y_i$ is a Gaussian measure at locations $(lat_i, lon_i)$, $i=1,...,n$, the model is 
\[y_i = \mu(lat_i,lon_i) +  \epsilon_i \quad ; \quad \epsilon_i \sim \mathcal{N}(0, \tau_{\epsilon}^{-1}),\] 
where $\mu(lat_i,lon_i)$ is a two-dimensional function, with no parametric assumptions on it and $\tau_{\epsilon}$ is the noise precision. We can think of $\mu(lat_i,lon_i)$ as a smooth surface representing the latent field which is modelled as a linear combination of bivariate B-spline basis functions: 
\[\mu(lat_i, lon_i) = \sum_{q=1}^Q \sum_{l=1}^L b_l(lat_i) b_q(lon_i) \beta_{l,q},\] where $b_l(lat_i) b_q(lon_i)$ is the tensor product of marginal B-splines, evaluated at $(lat_i,lon_i)$, and $\beta_{l,q}$ is the associated spline coefficient. The marginal B-splines $b_l, l=1,...,L$ ($b_q, q=1,...,Q$), are defined on a set of knots that are chosen to be equally-spaced over the latitude (longitude) domain. Taking the tensor product of the two marginal basis returns $K=QL$ bivariate B-splines built on a regular grid over the plane; see Figure \ref{fig:basis}, left panel. In this sense, P-splines give a low-rank representation of the latent field, as $K$ is typically chosen to be much lower than $n$. In matrix notation, $\bm \mu = \bm B \bm \beta$, where $\bm B$ is a basis matrix of dimension $n \times K$ and $\bm \beta$ the vector of spline coefficients. When data are organized in a regular grid with no missing values, the basis matrix can be computed by the Kronecker product $\bm B = \bm B_{lat} \otimes \bm B_{lon}$. When data are irregularly scattered over the plane, efficient row-wise Kronecker operations can still be used to compute $\bm B$, as this is equivalent to having data organized on a fine regular grid with missing values. We link the reader to \cite{eilers-2006} for details on P-splines for spatial data and to \cite{lee-thesis} for insights into the mixed model formulation of P-splines within a spatio-temporal setting.

P-splines have been framed in a fully hierarchical Bayesian context by \cite{brezger-2004}. The hierarchical model can be cast starting from the following likelihood:
\begin{eqnarray*}
\label{eq:lik}
\bm y | \alpha, \bm \beta, \tau_{\epsilon} & \sim & \mathcal{N}(\bm \mu,  \tau_{\epsilon}^{-1} \bm I) \quad ; \quad \bm \mu = \alpha + \bm B \bm \beta 
\end{eqnarray*}
The penalty is reproduced by an $r^{th}$ order random walk (RW) prior on the spline coefficients, that in general can be expressed as
\begin{equation}
\pi(\bm \beta|\tau_{\beta}) = (2\pi)^{-\texttt{rank}(\bm R)/2} (|\tau_{\beta} \bm R|^{*})^{1/2} \exp\left\{-\frac{\tau_{\beta}}{2}\bm \beta^{\textsf{T}} \bm R \bm \beta\right\},
\label{eq:igmrf}
\end{equation}
where $\tau_{\beta}$ is a scalar precision hyper-parameter and $\bm R$ is the structure matrix of dimension $K \times K$. The non-zero entries in $\bm R$ impose conditional dependencies among the spline coefficients, thus encoding the type of penalty. 
% The generalized determinant $|\tau_{\beta} \bm R|^{*}$ can be calculated as the product of the positive eigenvalues of $\tau_{\beta} \bm R$.
Formally, the RW is a particular type of Intrinsic Gaussian Markov Random Field (IGMRF). The smoothing properties of an IGMRF are determined by the pattern of non-zero entries of $\bm R$ and by its rank deficiency. Any vector in the null space of $\bm R$ can be added to $\bm \beta$ and density (\ref{eq:igmrf}) remains unchanged. For this reason, IGMRF priors are appropriate to model local deviations around an overall mean or, in general, a polynomial trend, with $\tau_{\beta}$ controlling the size of such deviations. For spatial smoothing, we will focus on a prior that leaves unspecified the overall mean, therefore  $\texttt{rank} (\bm R)=K-1$.

The precision matrix for P-spline smoothing over a plane proposed in \cite{eilers-2006} is constructed as the Kronecker sum
\begin{equation}
\label{eq:ksum}
\bm R = (\bm I_L \otimes \bm R_{lon}) + (\bm R_{lat} \otimes \bm I_Q)
\end{equation}
where $\bm R_{lat}$  and $\bm R_{lon}$ are the (marginal) structure matrices of a RW on latitudinal and longitudinal knots, respectively. If we take $\bm R_{lat}$ and $\bm R_{lon}$ as the structure of a $1^{st}$ order RW, this is equivalent to assume an intrinsic Conditional Autoregressive (ICAR) model \citep{besag}, with structure
%For penalization of the integrated squared second derivative we can set $r=2$, this giving a full structure matrix $\bm R$ of rank $K-2$. 
\begin{equation}
R_{ij} =  \begin{cases}
k_i & i=j\\
-1 & i \sim j\\
0 & \text{otherwise},
\end{cases}
\label{eq:structure}
\end{equation}
where $k_i$ is the number of knots adjacent to the $i^{th}$ knot; e.g. $k_i=\{2,3,4\}$ according to whether $i$ is a knot on the vertex, the border, or the interior of the regular grid. Usually an ICAR prior is assumed on a set of $n$ random effects, one for each data location, but here the ICAR is on the spline coefficients. In this sense, the basis $\bm B$ allows the stochastic field on the $K$ spline coefficients to be expanded at a much larger number of locations like $n$. This strategy allows substantial reduction of the number of parameters to estimate. 
Choosing higher order random walks in each dimension is possible: this will yield an higher order IGMRF prior, having a structure matrix with larger rank-deficiency; e.g. taking a 2$^{nd}$ order RW on latitude and longitude returns an IGMRF that models deviations from a plane. For a discussion of the properties of IGMRFs and their applications see \cite{rue-2005}.

\subsection{P-splines on Geodesic Discrete Global Grid Systems}
\label{sec:GDGG}

\begin{figure}
\centerline{
\includegraphics[width=0.5\textwidth]{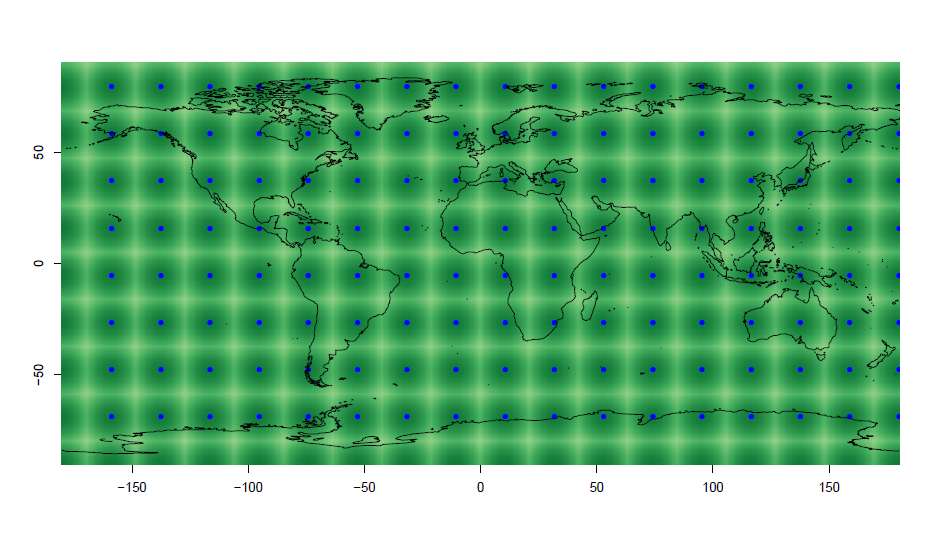}
\includegraphics[width=0.5\textwidth]{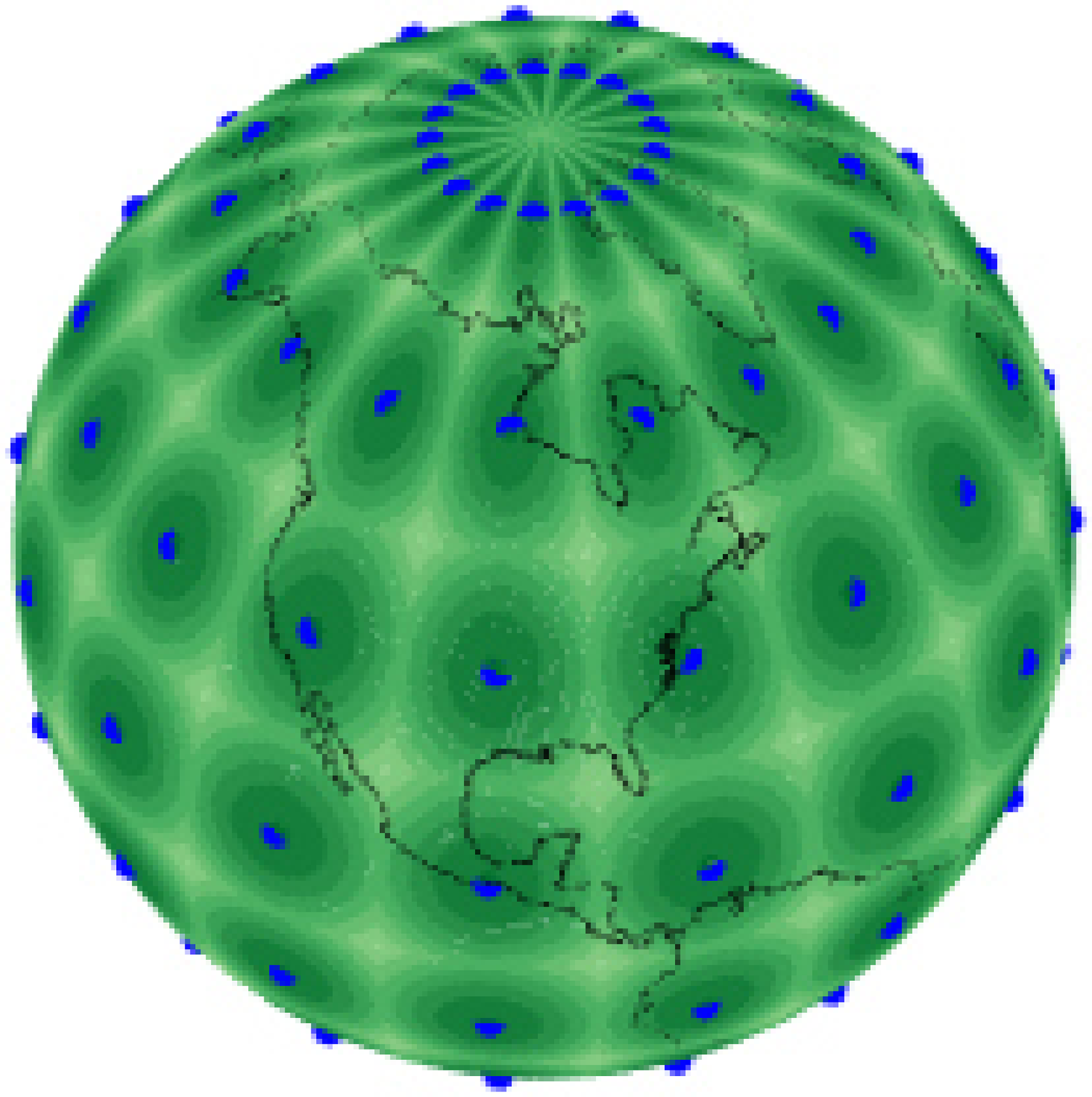}
}
\caption{Cubic B-splines equally-spaced in terms of Euclidean distances over the latitude and longitude plane (left panel; computed as the tensor product of marginal B-spline basis, see Section \ref{sec:classic-P-splines}). The right panel displays how these basis appear on the sphere. 
%\textit{...over-smoothing around equator because the basis are less compared to the poles...} 
}
\label{fig:basis}
\end{figure}

The assumption of equally-spaced knots is convenient for building Bayesian penalized spline models, because it allows to create a suitable smoothing prior by simply assuming an IGMRF model for regularly spaced locations on the spline coefficients. Following this idea, knot placement has to take into account the geometry of the support of the data. Thus, building an equally spaced basis on the latitude-longitude plane is not a sensible choice when data cover the whole Globe or a large region within it. Figure \ref{fig:basis} highlights that equally spaced B-splines in terms of Euclidean distances over the latitude-longitude plane (left panel) are not equally-spaced over the sphere (right panel). The spacing between the knots and the shape of the basis varies substantially latitude-wise: in such a knot-grid, imposing an IGMRF with structure (\ref{eq:structure}) and a single precision parameter $\tau_{\beta}$ on the spline coefficients would generate the spurious anisotropy discussed in \cite{banerjee-geo}. Of course, this would be a naive approach to spatial smoothing over the sphere, since it does not introduce conditional dependence between knots located at extreme longitudes, which are actually close on the sphere surface. A circular penalty imposing conditional correlations among these knots seems a more sensible choice, but the irregular knot placement over the sphere would still generate spurious non-stationarity, as will be discussed later on. In what follows we propose an approach for (a) building geodesic knot-grids which are quasi equally spaced in terms of geodesic distances, (b) building basis functions and penalty matrices on such grids. 

%%%% FIGURE basis

\subsubsection{Building the geodesic grid}
\begin{figure}
	\centerline{
		\includegraphics[width=0.33\textwidth]{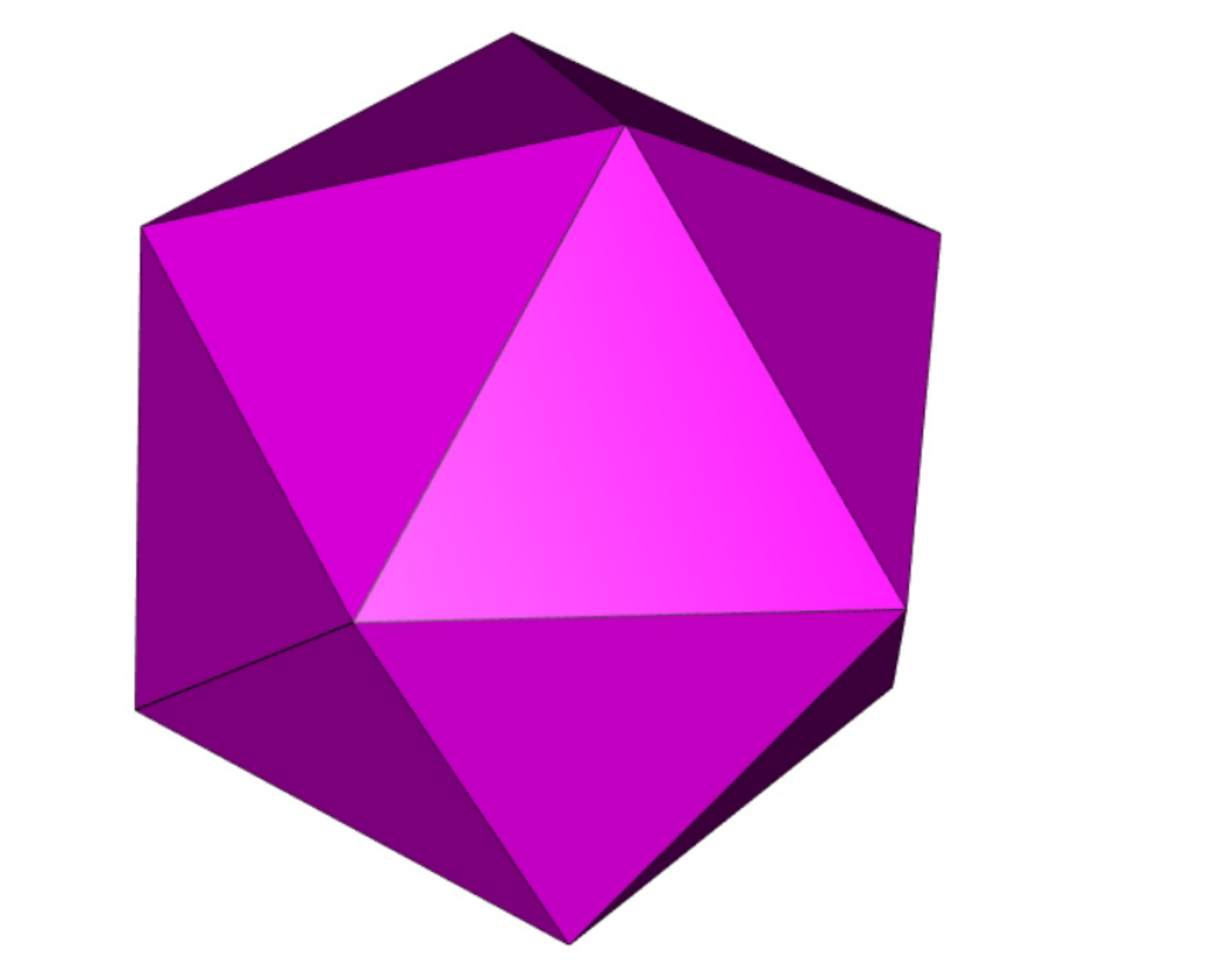}
		\includegraphics[width=0.33\textwidth]{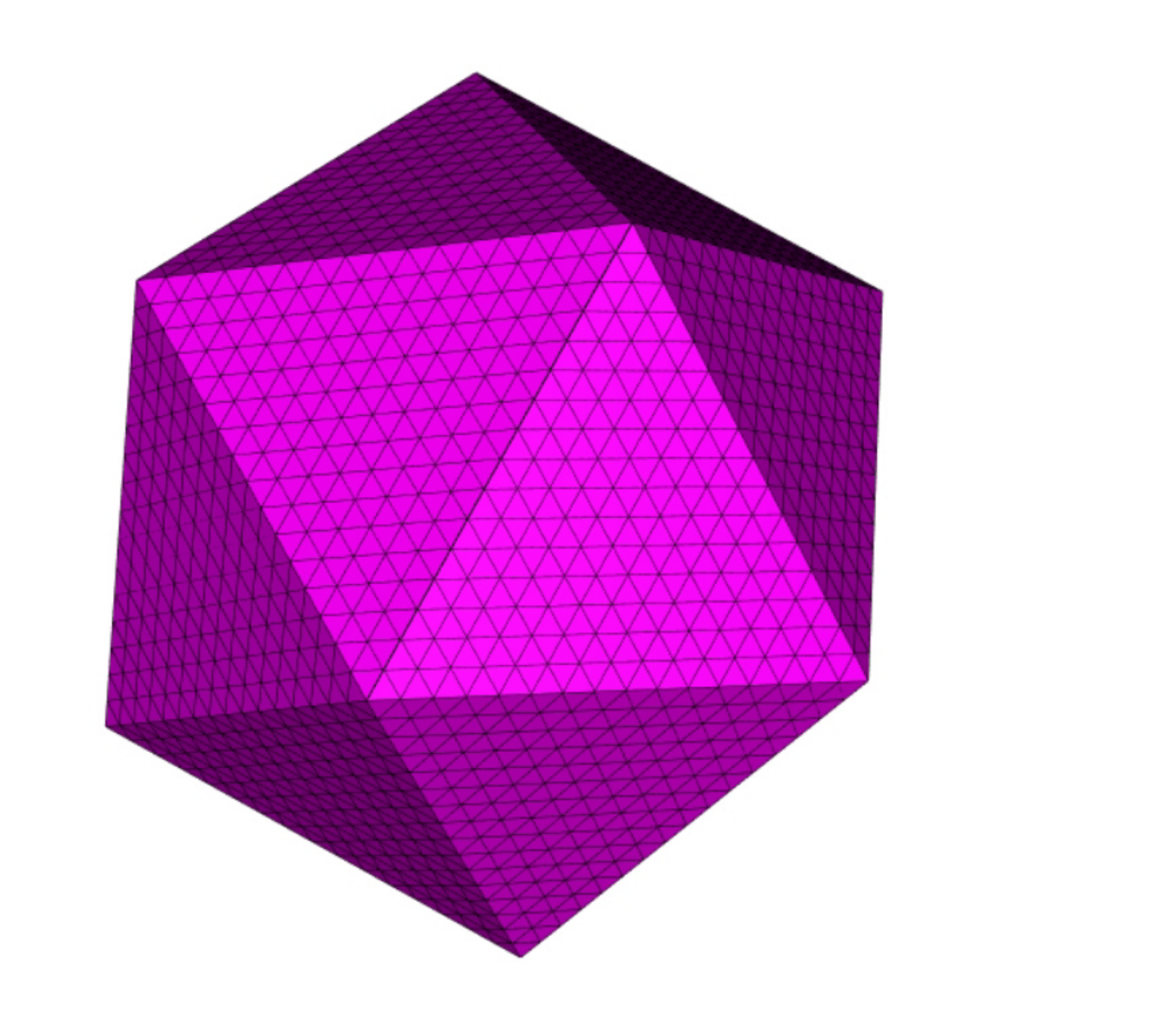}
		\includegraphics[width=0.33\textwidth]{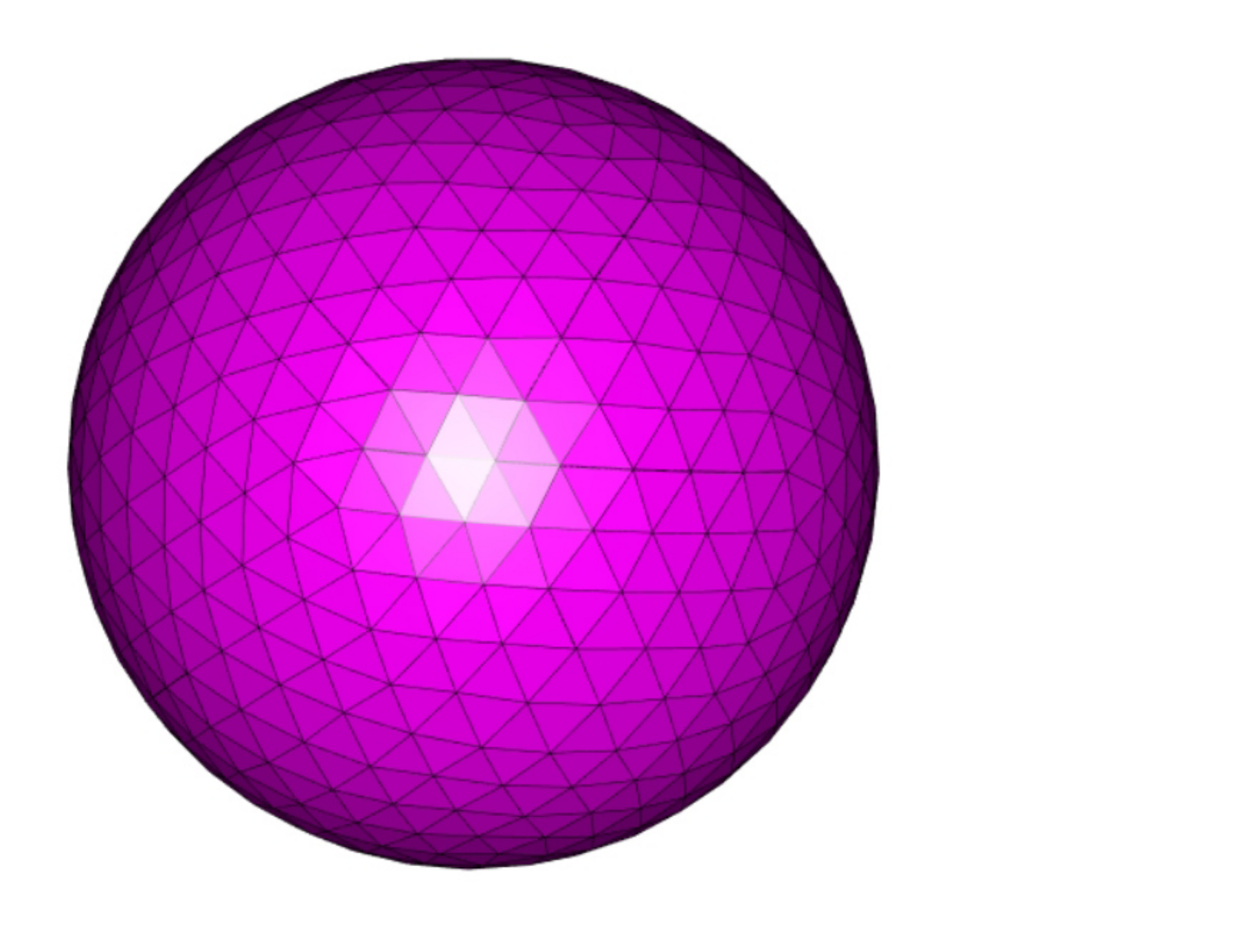}
	}
	\caption{On the left panel, the icosahedron. On the central panel, the icomesh, i.e. the regular triangular mesh after the \textit{split} operation is repeated four times ($\nu=4$). On the right panel, the icosphere, i.e. the mesh obtained from normalizing the icomesh nodes of the central panel.}% to have them laying on the sphere.}
	\label{fig:ico}
\end{figure}

Although building \textit{exactly} equally spaced grids over the sphere surface is an impossible task, GDGGs offer a close approximation to equal spacing and their architecture provides immediate solutions to build basis functions and penalty matrices. Details on the spatial configuration of GDGGs can be found in \cite{randall-2002}. \cite{sahr-2003} outline five design choices that need to be undertaken for GDGGs construction: our choices are listed below.
%listed below are our choices that returns a quasi regular triangular mesh over the sphere.
\begin{enumerate}
\item Choice of a \textit{base regular polyhedron}: we choose the {icosahedron}, which is a polyhedron made of 20 equilateral triangles and 12 nodes and assume this as a rough representation of a unit sphere. An icosahedron is displayed in Figure \ref{fig:ico}, left panel.
\item Choice of a fixed \emph{orientation} of the base regular polyhedron relative to the Earth: we set one node of the icosahedron at coordinates $(0,0,1)$, assuming this as the North Pole. 
\item Choice of a \textit{hierarchical spatial partitioning} method defined symmetrically on each face of the base regular polyhedron. At this step, we split each triangle of the icosahedron in four equal triangles. By repeating this operation an arbitrary number of times we obtain a refined mesh, which we denote as \emph{icomesh}. In Figure \ref{fig:ico}, central panel, see the icomesh resulting from four split iterations.
\item Transforming the base polyhedron partition into the corresponding spherical surface. This is achieved by simply normalizing the icomesh nodes, so that they lay on the sphere; we denote this mesh as \textit{icosphere}, see Figure \ref{fig:ico}, right panel. The icosphere is a refined icosahedron, hence a much better representation of the sphere. 
\item Choice of a method to \emph{assign points to grid cells}.  
Ability to assign point to grid cells composing the tessellation can be useful for several purposes. In our case, it is fundamental to determine which triangle a data location falls into when it comes to computation of the basis functions, as discussed in the following section. 
\end{enumerate}

Following the above five steps, we obtain a geodesic grid of knots which are quasi equally-spaced in terms of great-circle distances. To summarize, the GDGG is constructed by splitting each icosahedron face in four triangles, in a recursive way. Note that, while the icosphere is a sphere tessellated into spherical triangles, the icomesh is a regular mesh made by equilateral triangles. 

%%%% FIGURE icosahedron

\subsubsection{Building the basis and the penalty matrix}
\label{sec:bases}

The number of split iterations determines the dimension of the basis, i.e. the number of columns of the basis matrix $\bm B$ that we need to compute for each data point. Let $n$ be the number of data and $\nu$ be the number of split iterations, the basis has dimension  $n \text{ x } K$, with $K=5 \cdot 2^{2(\nu-1)+3} + 2$. We adopt B-spline basis functions centred at the knots, each basis spanning six triangles (thus assuming the six closest nodes as neighbours) except for those centred at the 12 icosahedron vertices (that have five neighbours). 

The next step consists of evaluating the $K$ B-splines, of a certain degree $d$, at an arbitrary data point laying on the sphere. Once that the triangle containing such point is determined, B-splines can be evaluated using Bernstein polynomials \citep{lai-2007spline}. To this aim, we find convenient working on the icomesh instead of the icosphere, as it is simpler to deal with planar than with spherical triangles. Therefore, we first project the 3d data location from the icosphere onto the icomesh domain, obtaining a point, $\bm v$, that falls inside a planar triangle (that lies on one of the icosahedron faces) and, second, we evaluate the $K$ B-splines at this 2d point. 
%(Note that $\bm v$ can be represented by 2-dimensional coordinates as it lies on the plane of the corresponding icosahedron face.) 
Following \cite{lai-2007spline}, any point $\bm v = (x,y)$ inside a triangle of vertices $\bm v_1=(x_1, y_1), \bm v_2=(x_2, y_2), \bm v_3=(x_3,y_3)$ has a unique representation as 
 
\begin{equation*}
\bm v=\bm v_1 b_1 + \bm v_2 b_2 + \bm v_3 b_3,
\end{equation*}
where $(b_1, b_2, b_3)$ are called barycentric coordinates and are such that $b_1+b_2+b_3=1$. 
%The barycentric coordinates $(b_1,b_2,b_3)$ are associated to the first, second and third vertices of the triangle. 
The Bernstein polynomial of degree $d$ is 
\begin{equation}
H_{tjk}^d = \frac{d!}{t!j!k!}b_1^t b_2^j b_3^k
\label{eq:bernstein}
\end{equation}
with $t,j,k$ integer numbers summing to $d$. The following property
\begin{equation*}
\sum_{t+j+k=d} H_{tjk}^d =1
\end{equation*}
guarantees that for each location on the sphere the basis functions sum to 1. This is a desirable property of any smoothing model, giving a flat spatial field when there is no variation around the overall level, i.e. all spline coefficients are equal. 

Let us indicate with $\bm z_i=(z_{i1},z_{i2})$ the location for observation $i$ projected on the icomesh, with $\bm B[i,]$ the row entry of $\bm B$ with the B-splines evaluated at $\bm z_i$ and with $\{k_1, k_2,k_3\}$ the indexes for the three knots closest to $\bm z_i$ (note, these are the vertices of the triangle containing observation $i$). It is important to note that only the B-splines centered at $\{k_1, k_2,k_3\}$ are non-zero at $\bm z_i$, whereas the B-splines centered at the remaining knots in the icomesh are zero at $\bm z_i$.
The three non zero element of $\bm B[i, \{k_1, k_2,k_3\}]$ can be expressed as Bernstein polynomials (\ref{eq:bernstein}), i.e. polynomials in the barycentric coordinates. Table \ref{tab:bernstein} reports the non zero elements of $B[i,]$ for linear ($d=1$), quadratic ($d=2$) and cubic ($d=3$) B-splines.
%The three non zero element of $\bm B[i, \{k_1, k_2,k_3\}]$ are given below, considering a linear ($d=1$), quadratic ($d=2$) and cubic ($d=3$) order,
%\begin{itemize}
%\item  $d=1: \bm B[i, \{k_1, k_2,k_3\}] = (H_{100},H_{010},H_{001}) = (b_1,b_2,b_3)$;
%\item $d=2: \bm B[i, \{k_1, k_2,k_3\}] = (H_{200}+H_{100},H_{020}H_{010},H_{002}+H_{001})=(b_1^2+b_1,b_2^2+b_2,b_3^2+b_3)$;
%\item $d=3: \bm B[i, \{k_1, k_2,k_3\}] = (H_{300}+H_{200}+H_{100},H_{030}+H_{020}H_{010},H_{003}+H_{002}+H_{001}) =(b_1^3+b_1^2+b_1,b_2^3+b_2^2+b_2,b_3^3+b_3^2+b_3)$,
%\end{itemize}

\begin{table}
\small
\centering
\begin{tabular}{|c|c|c|}
\hline
$d=1$ & $d=2$ & $d=3$ \\
\hline\hline
$ 
\left(\begin{array}{c}
H^1_{100} \\ H^1_{010} \\ H^1_{001} 
\end{array}
\right)=
\left(\begin{array}{c}
b_1 \\ b_2 \\ b_3
\end{array}
\right)
$
& 
$ 
\left(\begin{array}{c}
H^2_{200}+H^2_{100} \\ H^2_{020}+H^2_{010} \\ H^2_{002}+H^2_{001} 
\end{array}
\right)=
\left(\begin{array}{c}
b_1^2+b_1 \\ b_2^2+b_2 \\ b_3^2+b_3
\end{array}
\right)
$
& 
$ 
\left(\begin{array}{c}
H^3_{300}+H^3_{200}+H^3_{100}\\ H^3_{030}+H^3_{020}+H^3_{010} \\ H^3_{003}+H^3_{002}+H^3_{001}
\end{array}
\right)=
\left(\begin{array}{c}
b_1^3+b_1^2+b_1 \\ b_2^3+b_2^2+b_2 \\ b_3^3+b_3^2+b_3
\end{array}
\right)
$
\\ 
\hline 
\end{tabular}
\caption{Non zero elements of $\bm B[i,]$, for B-splines of degree $d=\{1,2,3\}$.}
\label{tab:bernstein}
\end{table}

\begin{figure}
	\centerline{
		\includegraphics[width=0.5\textwidth]{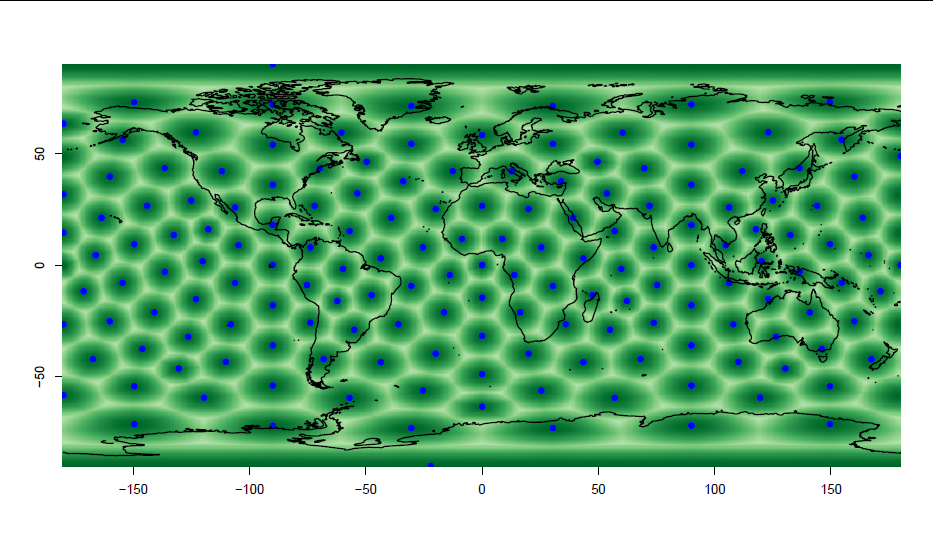}
		\includegraphics[width=0.5\textwidth]{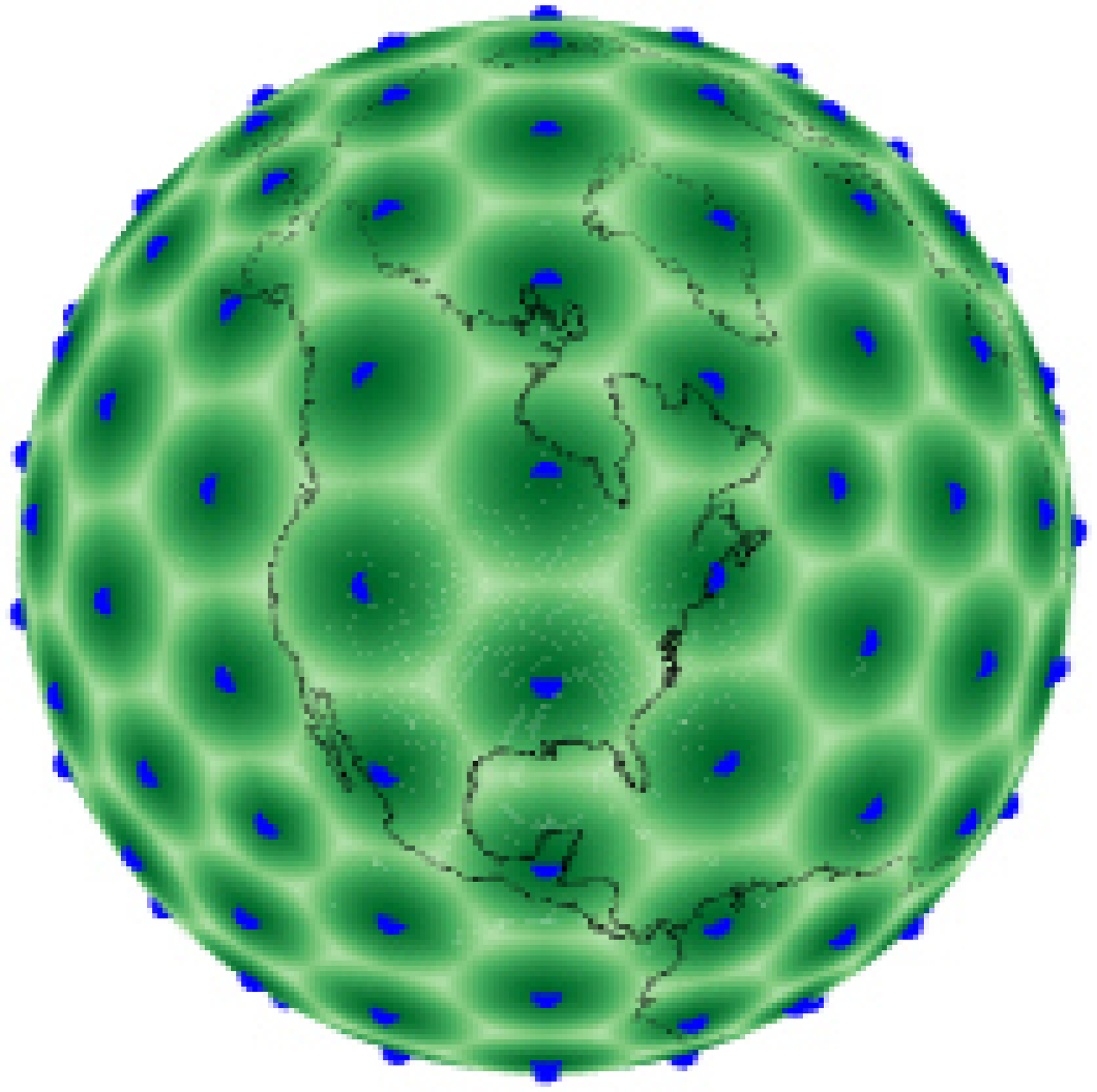}
	}
	\caption{Cubic B-splines equally-spaced in terms of geodesic distances over the sphere (right panel; computed using Bernstein polynomials on a GDGG, see Section \ref{sec:GDGG}). The left panel displays how these basis appear on the latitude longitude plane.}
	%\caption{
	%Left hand panels: cubic B-splines equally-spaced in terms of euclidean distances. Right hand panels: cubic B-splines equally-spaced in terms of geodesic distances.}
	%
	%(We chose a small number of B-splines for illustrative purposes, in practical applications the number of basis functions has to be bigger).
	\label{fig:gbasis}
\end{figure}

The resulting basis matrix $\bm B$ is sparse because the B-splines are non zero over a domain spanned by only six triangles on the icomesh. Figure \ref{fig:gbasis}, left panel, shows how the new basis functions appear when projected over latitude and longitude.  This plot suggests that a fairly similar degree of smoothness is applied everywhere using this new basis, avoiding the kind of spurious anisotropy introduced by the basis in Figure \ref{fig:basis}. 
The Geodesic P-splines setting is completed by specifying the matrix $\bm R$, that we choose as the ICAR structure (\ref{eq:structure}) with rank-deficiency 1. The number of neighbouring knots is $k_i=5$, if $i$ is one of the 12 nodes of the icosahedron, and $k_i=6$, if $i$ is one of the remaining $K-12$ nodes.

It is worth noting that, when using IGMRF priors with precision matrix $\tau_{\beta} \bm R$ on the spline coefficients $\bm \beta$, the structure of conditional dependence imposed by $\bm R$ determines the structure of the marginal variances of each coefficient, $Var(\beta_i)=\tau_{\beta}^{-1} R^{-}_{ii}, i=1,\ldots,K$, $\bm R^-$ being the generalised inverse of $\bm R$. Different structures can lead to extremely different marginal variances. To account for this feature, \cite{sorbye-2013} suggest to scale the precision matrix so that the hyperprior for $\tau_{\beta}$ can be selected to give the same degree of smoothness, a priori, starting from different structure matrices. 
%This scaling can be achieved by multiplying $\bm R$ for the geometric mean of the marginal variances, so that the average marginal variance of the scaled model is equal to 1. 
The scaled precision matrix can be obtained as $\bm R^*=\kappa \bm R$, where $\kappa$ is the geometric mean of the diagonal entries of $\bm R^-$. IGMRFs with scaled precision matrices, although being characterised by different correlation structure, have a common feature: the average marginal variance is equal to one.

\begin{figure}
	\centerline{
		\includegraphics[width=6cm]{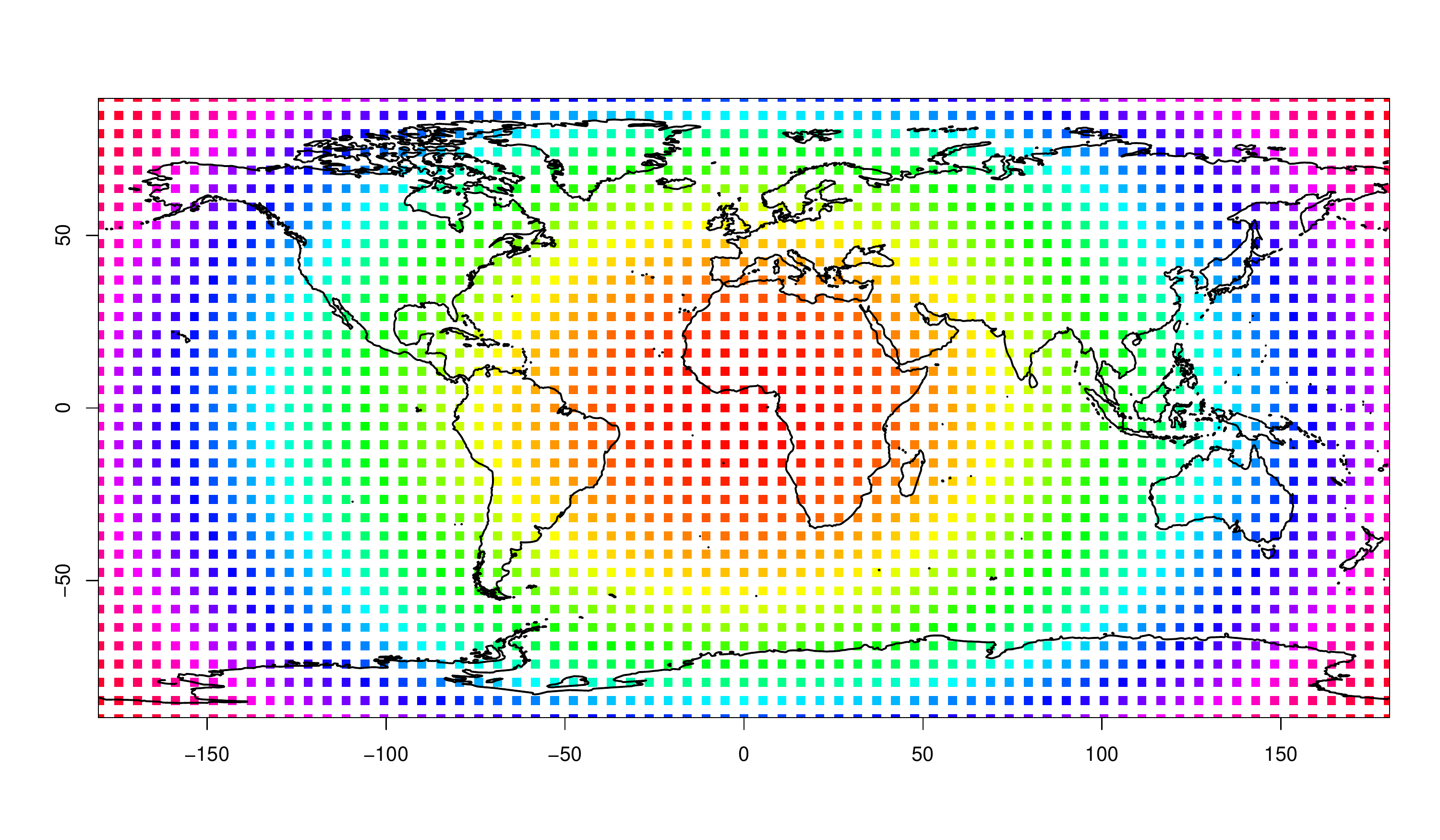}
		\includegraphics[width=6cm]{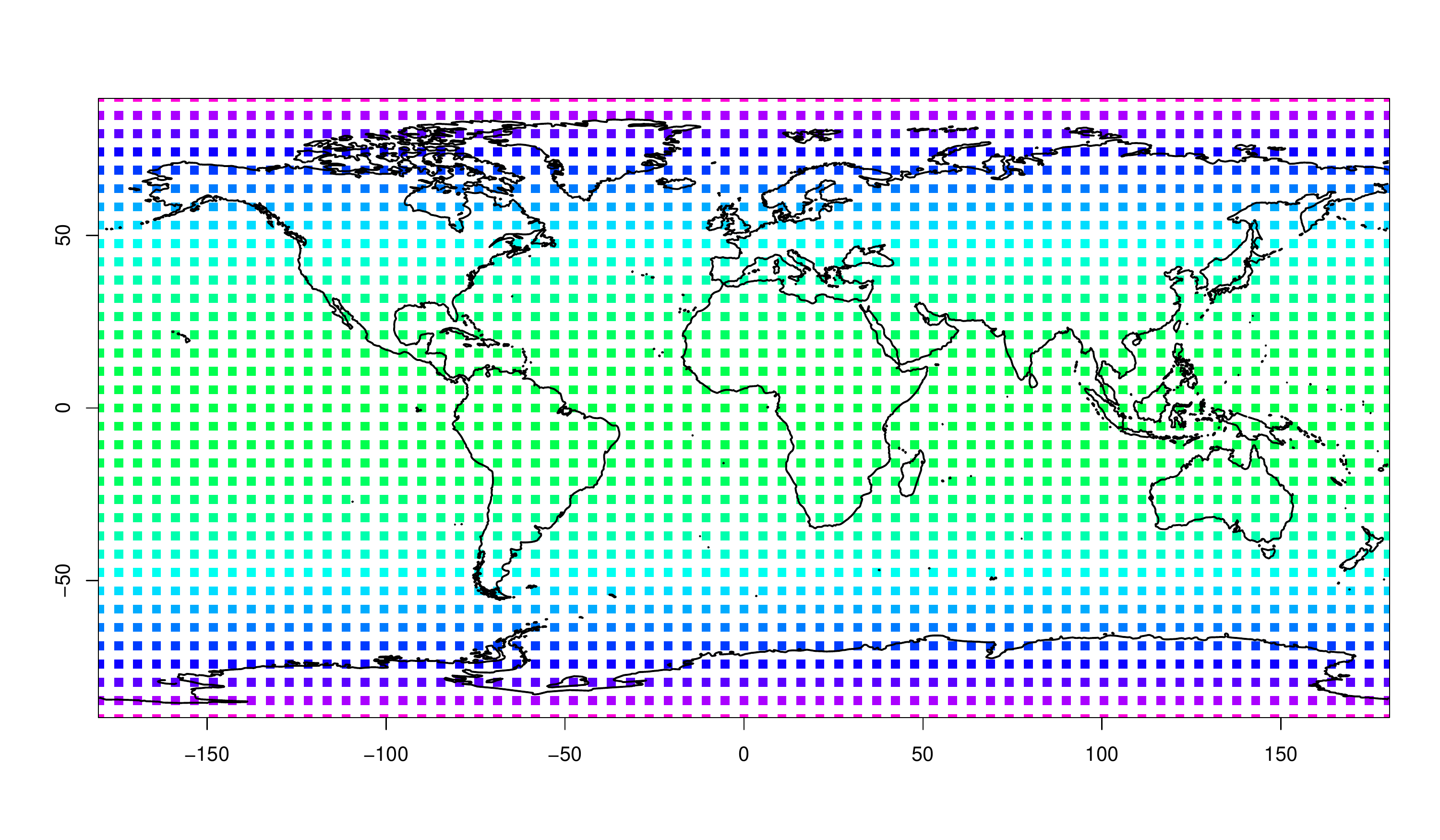}
		\includegraphics[width=.65cm]{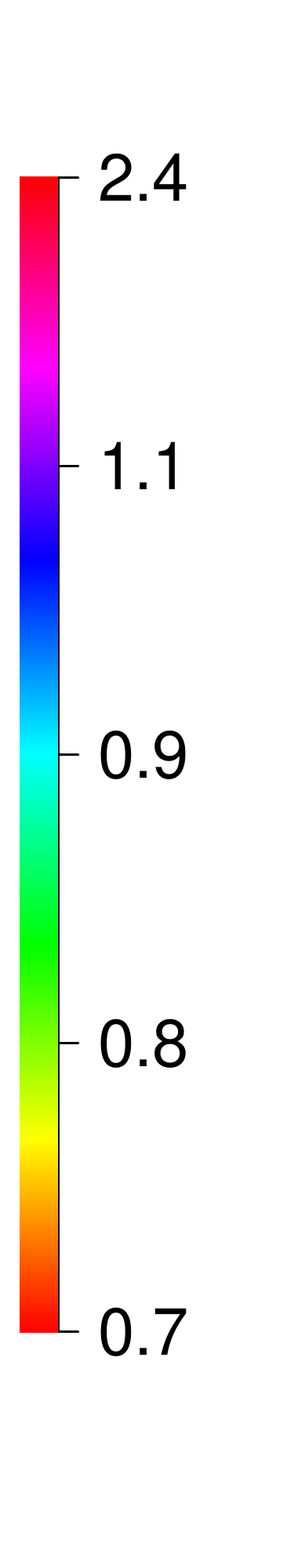}
				}
		\centerline{
		\includegraphics[width=6cm]{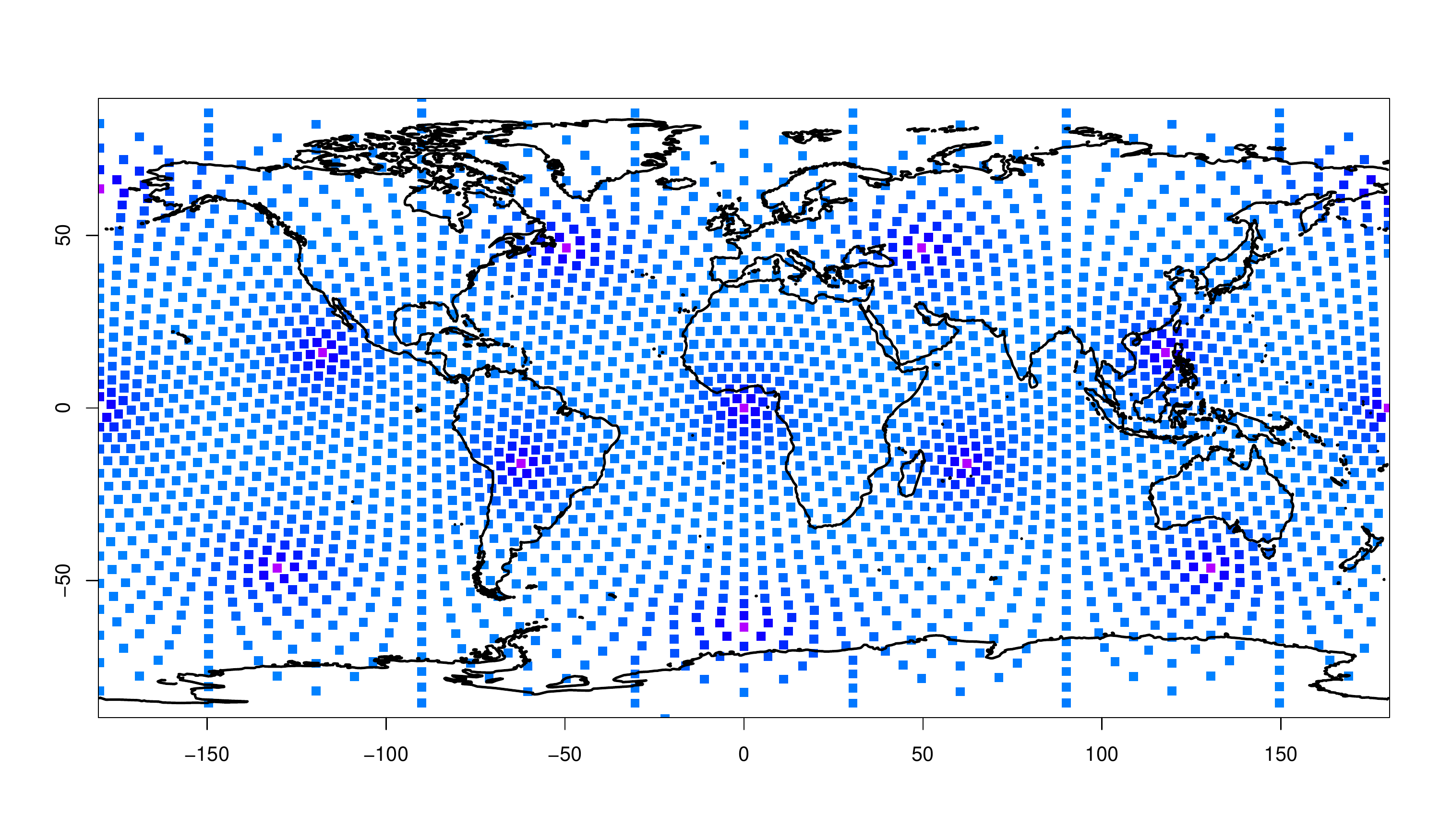}
	}
\caption{
Marginal variances with naive penalty (left panel), longitude-wise circular penalty (central panel) and geodesic penalty (right panel). (The color bar on the right is valid for all the three panels). 
}
\label{fig:mvar}
\end{figure}

Figure \ref{fig:mvar} compares the marginal variances for three models corresponding to a naive penalty (left), longitude-wise circular penalty (central) and a geodesic penalty (right). For the sake of comparison the precision matrices associated with the three models were scaled. For naive penalty, we mean an IGMRF prior for the spline coefficients laying on a planar grid, using the ICAR structure (\ref{eq:structure}). The longitude-wise circular penalty is an IGMRF on a planar grid with structure (\ref{eq:ksum}), but assuming $\bm R_{lon}$ as the structure of a circular $1^{st}$ order RW. For geodesic penalty, we mean an IGMRF on a GDGG using the ICAR structure as described in Section \ref{sec:bases}. In the left panel, the non-stationarity in the marginal variances implied by using the ICAR structure on a regular planar grid (naive penalty) is evident. In the middle panel, marginal variances obtained by building a circular penalty longitude-wise show variation latitude-wise as expected. The IGMRF prior on the geodesic grid with ICAR structure implies stability in the marginal variances that is not achieved with the other specifications. %We believe it is a good knot placement and penalisation strategy for data on the sphere. 
As a matter of fact, the geodesic grid is almost a torus since all knots, except the icosahedron nodes, have six neighbours; we believe this is a desirable feature of our model as it mimics the idea of stationarity in variance typical of Mat\'{e}rn correlation functions. 

%[PEZZO VECCHIO non spiegava tutto...]\textit{In Figure \ref{fig:mvar}, left panel, we display the marginal variances at the knot of the planar grid, after scaling the precision matrix (\ref{eq:ksum}): the non-stationarity in the marginal variances implied by this structure is evident. In the middle panel, we consider marginal variances obtained by building a circular penalty longitude-wise (\cite{rue-2005}), while right panel shows marginal variances obtained with the geodesic grid.}

%%%% FIGURE Mvar

\subsubsection{Hyperpriors}
To complete the fully Bayesian model we need to set priors for the hyper-parameters $\tau_{\beta}$ and $\tau_{\epsilon}$. The precision $\tau_{\beta}$ regulates the amount of smoothing.  When $\tau_{\beta}$ goes to infinity, $\bm \mu$ is a constant (because the rank deficiency of $\bm R$ is 1), while $\tau_{\beta} \in (0,+\infty)$ give a more flexible surface. A standard approach is to use a Gamma, $\text{Ga}(a,b)$, with shape $a$ and rate $b$, for both random walk and noise precisions. Usual parametrizations are $a$ equal to $1$ and $b$ small (e.g. $\text{Ga}(1,5e-5)$), or $a$ and $b$ small (e.g. Gamma$(1e-3,1e-3)$), as an attempt of non informativeness on the variance scale. Several papers in the literature have discussed issues related to the Gamma conjugate priors in hierarchical additive models and proposed alternatives \citep{gelman-2006, pcprior}. Typically, the main impact regards the prior for the random walk precision, whereas the prior for the noise precision is negligible. In general, choice about the prior $\pi(\tau_{\beta})$ will be relevant in situations where we have poor sample size compared to the number of parameters to estimate. In the case study under examination the large sample size available for estimating each spline coefficient makes the impact of $\pi(\tau_{\beta})$ very small.

\subsubsection{Computations}
Model estimation does not raise particular issues with respect to planar P-spline models, once matrices $\bm B$ and $\bm R$ have been built. Indeed, the model belongs to the class of Latent Gaussian Markov Models and approximate Bayesian inference can be performed efficiently using the package \texttt{R-INLA} \citep{rue-inla}. In our case study, we find more appropriate to use a Gibbs sampling algorithm as the tools developed for detecting the ITCZ require a sample from the joint posterior distribution of the model. 
%, a task that we found more practical to perform by keep on sampling from the MCMC chain of the parameters of interest after convergence of the MCMC algorithm. 
The most expensive step is to sample from the full conditional for the spline coefficients 
\begin{equation}
\bm \beta|\tau_{\beta}, \tau_{\epsilon},\bm y \sim  N(\bm Q^{-1} \bm B^{\textsf{T}} \bm y,  \bm Q^{-1}) \quad \quad \quad  \bm Q= \left(\bm B^{\textsf{T}} \bm B + {\frac{\tau_{\beta}}{\tau_{\epsilon}}} \bm R \right) 
\label{eq:fullcond_beta}
\end{equation}
under the linear constraint $\bm 1_{K}^{\textsf{T}} \bm B \bm \beta = 0$ needed for intercept identifiability. We use an efficient Gibbs sampler coded in \texttt{R} with the use of sparse matrix algebra as implemented in the \texttt{spam} package \citep{furrer-2010} to exploit sparsity of $\bm Q$ in (\ref{eq:fullcond_beta}). The \texttt{spam} package contains routines to perform efficient Cholesky decomposition of $\bm Q$, which is important for fast sampling from a GMRF under linear constraints like the full conditional $\pi(\bm \beta|\tau_{\beta}, \tau_{\epsilon},\bm y)$ in (\ref{eq:fullcond_beta}). The full conditionals for all the parameters in the model and the code for implementing the Gibbs sampler in \texttt{R} can be found in the supplementary material. 

%\begin{table}[ht]
%\centering
%\begin{tabular}{c|ccccccc}
%  \hline
%num.splits & 0 & 1 & 2 & 3 & 4 & 5 & 6\\
%num.nodes  & 12 & 42 & 162 & 642 & 2562 & 10242 & 40962\\ 
%   \hline
%\end{tabular}
%\label{tab:split}
%\end{table}

%%%% FIGURE B-splines
%\begin{figure}
%\centerline{
%\includegraphics[scale=0.5]{fig/bivariate-b-spline-over-tri-d1d3.pdf}
%}
%\vspace{-2cm}
%\caption{On the left panel ($d=1$): the B-spline is equal to 1 in the central node and decays (linearly) to 0 in the six neighbouring nodes. On the right panel ($d=3$): the B-spline is equal to 1 in the central node and decays (as a polynomial of degree 3) to 0 in the six neighbouring nodes. In each case, the B-spline is centred at the central node and spans a region formed by six grid cells around it.}
%\label{fig:bsplines}
%\end{figure}

%\input{sec-results2}
\section{Application}
\label{sec:results}
\subsection{Modelling TCWV data}
\label{sec:smoothing-TCW}

The goal of our application is to detect the ITCZ location by using the TCWV dataset described in Section \ref{sec:motivating}. The operative definition of ITCZ that we use, as suggested by researchers from ISAC-CNR, Italy, is ``the strip surrounding the Earth surface where TCWV shows highest values''.\\
To this aim, we first apply Geodesic P-splines for smoothing observed TCWV data, which are affected by noise and do not provide measurements over the land, in order to predict the latent field all over the world. Then, we exploit model output for locating ITCZ by sampling form the joint posterior distribution of the latent field.
Let $\bm y=(y_1,\dots,y_n)^{\textsf{T}}$ be the vector of TCWV observations $i=1,\ldots,n$, the hierarchical model is 
\begin{eqnarray}
\text{Likelihood:} \nonumber\\
\bm y | \alpha, \bm \beta, \tau_{\epsilon} & \sim & \mathcal{N}(\bm \mu,  \tau_{\epsilon}^{-1} \bm I) \nonumber\\
\bm \mu & = & \alpha + \bm B \bm \beta \label{eq:mu}\\
\text{Prior:}  \nonumber\\
\alpha & \sim & \mathcal{N}(0, \tau_{\alpha}^{-1})  \nonumber \\
\bm \beta | \tau_{\beta} & \sim & \mathcal{N}(\bm 0, \tau_{\beta}^{-1} \bm R^{*})  \quad \quad \text{$\bm \beta$ is subject to $\bm 1_{K}^{\textsf{T}} \bm B \bm \beta = 0$}
\label{eq:constraint}  \\
\text{Hyper-prior:} \nonumber\\
\tau_{\beta} & \sim & \text{Ga}(1,5e-5)\nonumber \\
\tau_{\epsilon} & \sim & \text{Ga}(1,5e-5)\nonumber 
\end{eqnarray}

At the likelihood level, the matrix $\bm B$ in (\ref{eq:mu}) is the B-spline basis on a GDGG as described in Section \ref{sec:GDGG}. The latent field $\bm \mu$ is a surface varying smoothly over the sphere, with $\alpha$ the global spatial mean and $\bm \beta$ the spline coefficients. At the prior level, we have a diffuse Gaussian prior, with $\tau_{\alpha}$ fixed at a small value for the intercept and an ICAR prior, with precision $\tau_{\beta} \bm R^*$, for the spline coefficients. Using the scaled matrix $\bm R^*$ is a fundamental step: this allows to select the same prior for $\tau_{\beta}$ and $\tau_{\epsilon}$, as both $\bm I$ in (\ref{eq:mu}) and $\bm R^{*}$ in (\ref{eq:constraint}) have average marginal variance equal to 1. The results presented in this section are obtained using a $\text{Ga}(a=1,b=5e-5)$ for both hyperparameters, after checking that the results were non sensitive to other choices for $a$ and $b$. 
%\textit{Note that for identifiability of the spline coefficients and the intercept, one linear constraint is needed as indicated in equation (\ref{eq:constraint}).} 

We fitted the Geodesic P-spline model to the data displayed in Figure \ref{fig:data}, referred to January and July, 2008. Data come as a raster of $720 \times 360$ observations geo-referenced in latitude longitude coordinates, with the percentage of raster cells with missing observations being about $40\%$ for both months.
%$163$ rows (or latitude points) and $360$ columns (or, longitude points) with more than $20000$ (exactly, $21074$) missing values in correspondence of the raster cells covering land. These observations are unavailable because the instrument cannot retrieve the TCWV measure in a reliable way on land. The only measures available on land are those in correspondence of lakes (see, for instance, the lake area in north America, Figure \ref{fig:data}). 
To compute $\bm B$, the latitude and longitude coordinates were converted to spherical coordinates, then projected on the icomesh and finally cubic B-splines were evaluated on a GDGG with $K=10242$ nodes (i.e. $\nu=5$). When using P-splines, the dimension of the knot grid has to be specified by the user according to the application at hand. It is required to select enough basis in order to capture the spatial variability of the latent field. As long as the basis is large enough to track the signal, increasing the number will not change the fit, rather it will change the location of posterior $\pi(\tau_{\beta}|\bm y)$ (analogously to the rescaling of the smoothing parameter when changing the number of knots). This means that a prior for $\tau_{\beta}$ has to be chosen with care in general but especially in poor sample size contexts, which is not the case here. 
Model estimation is performed by Gibbs sampling: we draw a total of $5000$ samples after convergence (achieved after a quick burnin due to the relative large sample size available for each model parameter). Regarding computational time, it takes about fifteen seconds (with an Intel core, 2.00GHz, 8.00 ram) to run a hundred iterations when $\nu=5$.

\begin{figure}
\centerline{
		\includegraphics[width=0.475\textwidth]{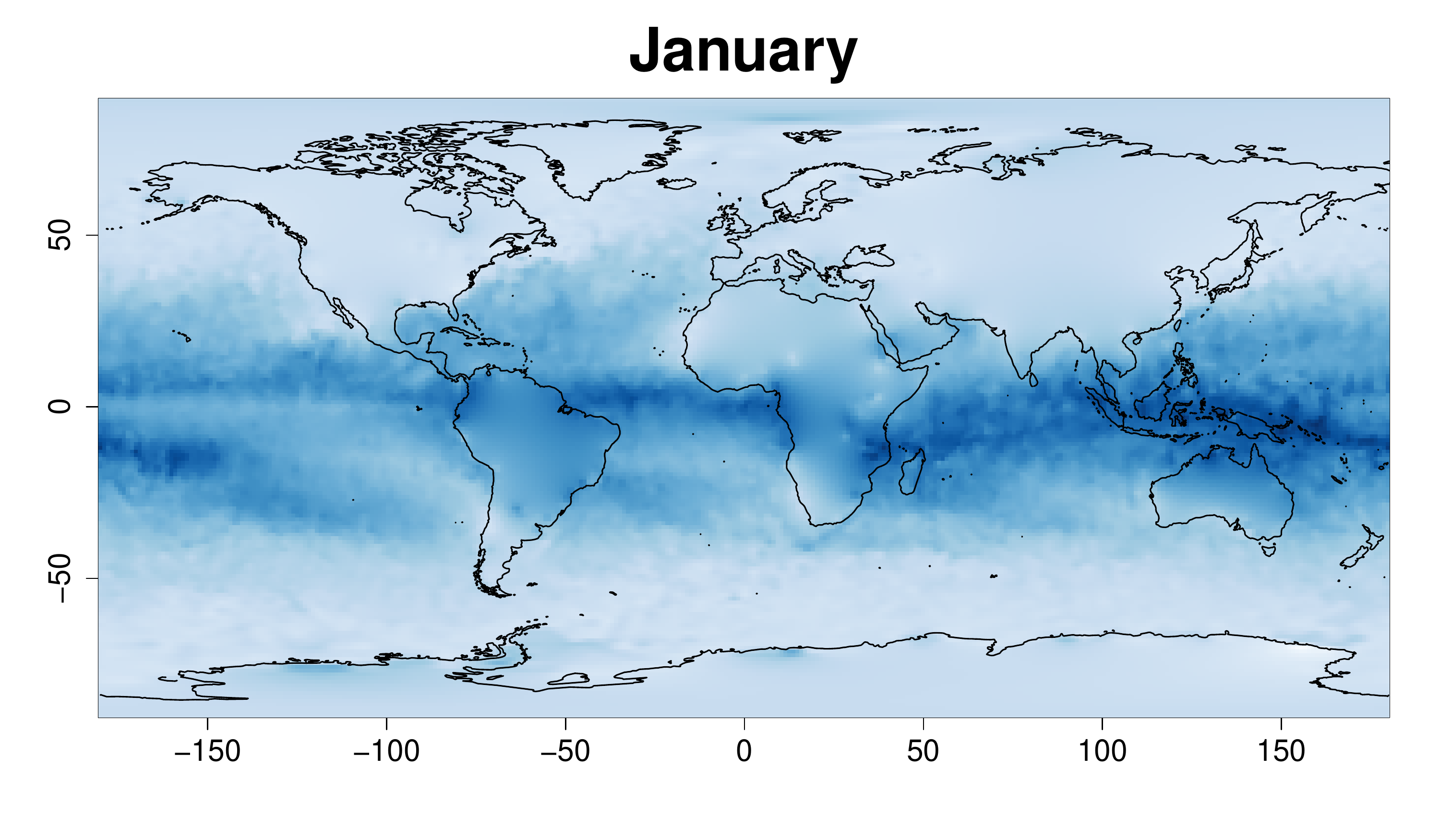}
		\includegraphics[width=0.475\textwidth]{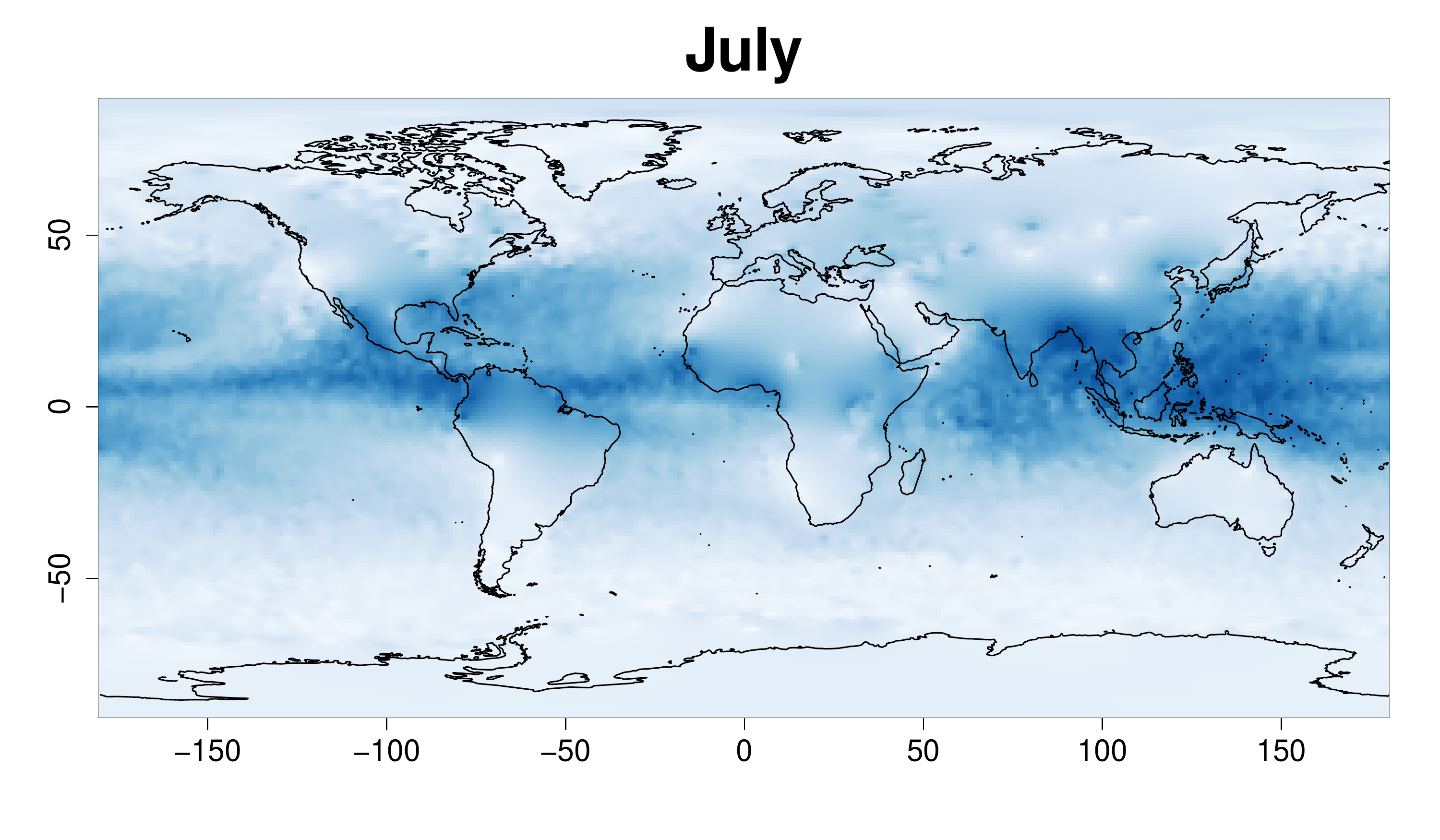}
	\includegraphics[width=.05\textwidth]{colorbar.pdf}
}
	\caption{Model prediction of the latent field for January and July.}
	\label{fig:smooth}
\end{figure}

In a Bayesian framework, spatial prediction is naturally based on the joint posterior predictive distribution of the latent field: sampling from this distribution is particularly efficient when using Bayesian P-splines. 
Once the posterior distribution of the latent field $\pi(\bm \mu|\bm y)$ has been obtained, prediction $\tilde \mu$ at an arbitrary location $\bm{\tilde x}$ can be performed, after evaluating the basis functions at $\bm{\tilde x}$, using the posterior predictive distribution:

\begin{equation}
\label{eq:pred}
\pi(\tilde \mu|\bm y) = \int \pi(\tilde \mu|\bm \theta)\pi(\bm \theta|\bm y)d \bm \theta
\end{equation}

where $\bm \theta=(\alpha,\bm \beta,\tau_\alpha,\tau_\beta,\tau_\epsilon)$. This is achieved by composite sampling once $G$ samples from the posterior distribution are available. Let $\bm \theta^g$ be a sample from the posterior distribution, $g=1,\ldots,G$: samples from distribution (\ref{eq:pred}) are obtained by sampling from $\pi(\tilde \mu|\bm \theta^g)$.
%The posterior predictive distribution of the latent field at arbitrary point $\tilde x$ can be achieved as $\tilde{\bm\mu} = \bm \tilde{B} \hat{\bm\beta}$, where matrix $\tilde{\bm B}$ contains B-splines evaluated at the prediction points (given the same GDGG used to fit the model) and $\hat{\bm \beta}$ is the posterior mean of the spline coefficients obtained via MCMC. 
In Figure \ref{fig:smooth} we report the maps of the TCWV posterior means at a fine grid of prediction locations: this accomplishes our first task, i.e. to remove random noise from data and to reconstruct the latent field on the whole Earth surface. Our strategy for ITCZ location is outlined in what follows.

\subsection{Locating the ITCZ}
\label{sec:locating-TCW}

The problem of ITCZ location is addressed by summarising the posterior predictive distribution of the TCWV latent field. The procedure outlined below requires the specification of a reasonable guess concerning the width of the ITCZ region denoted as $W$; we based our choice on expert knowledge by ISAC-CNR researchers and set $W=1000$ $km$. The ITCZ width relative to the length of a Meridian (which is about $20000$ $km$) is around $w=W/20,000=0,05$.

Our algorithm to locate the ITCZ consists of a discrete search performed longitude-wise (i.e. at each meridian). Let $m=1,\ldots,M$ index a set of $M$ meridians: for a given $m$, we sample from the posterior predictive distribution of the latent field at a fine grid over latitude. Then, we compute the posterior probability that a point at a given latitude belongs to the region where the TCWV shows highest values (i.e. the point falls into the ITCZ region), integrating out uncertainty about model parameters.

\begin{figure}
\centerline{
		\includegraphics[width=0.475\textwidth]{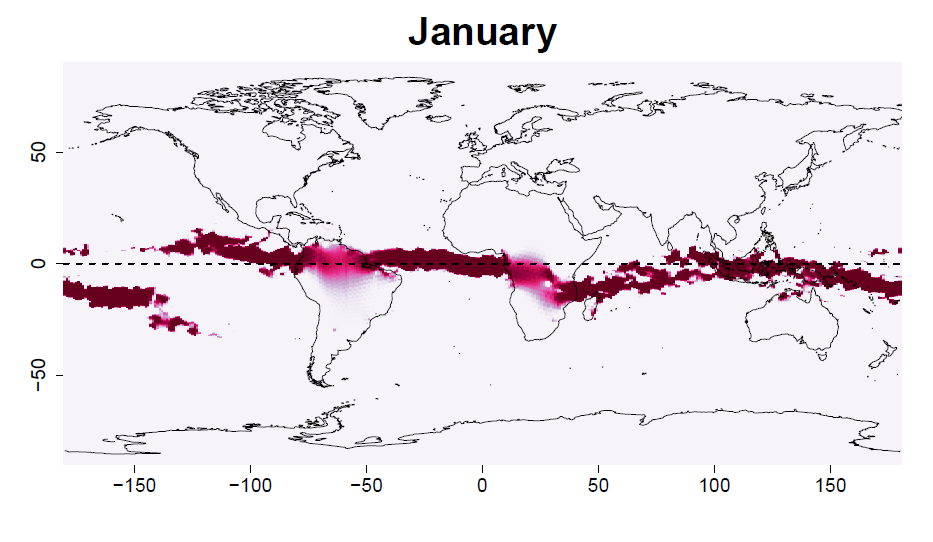}
		\includegraphics[width=0.475\textwidth]{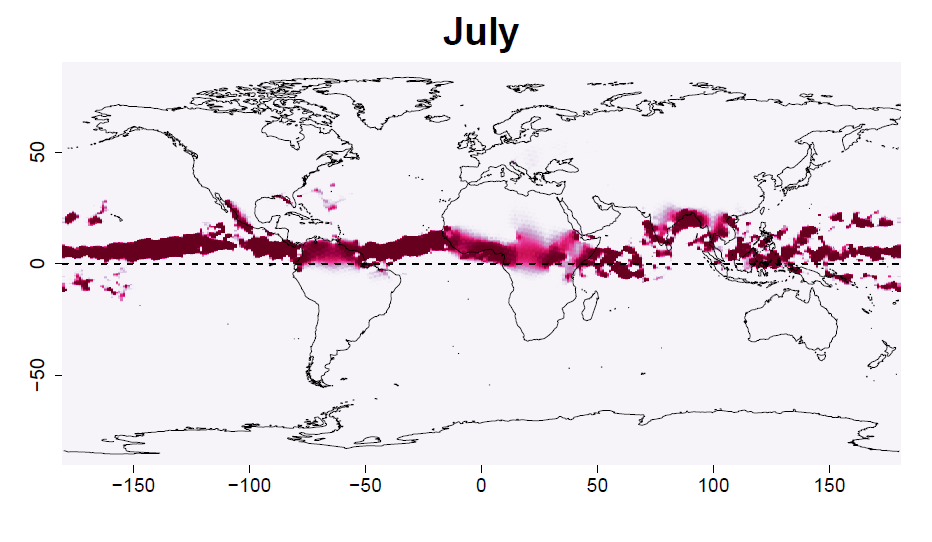}
		\includegraphics[width=.05\textwidth]{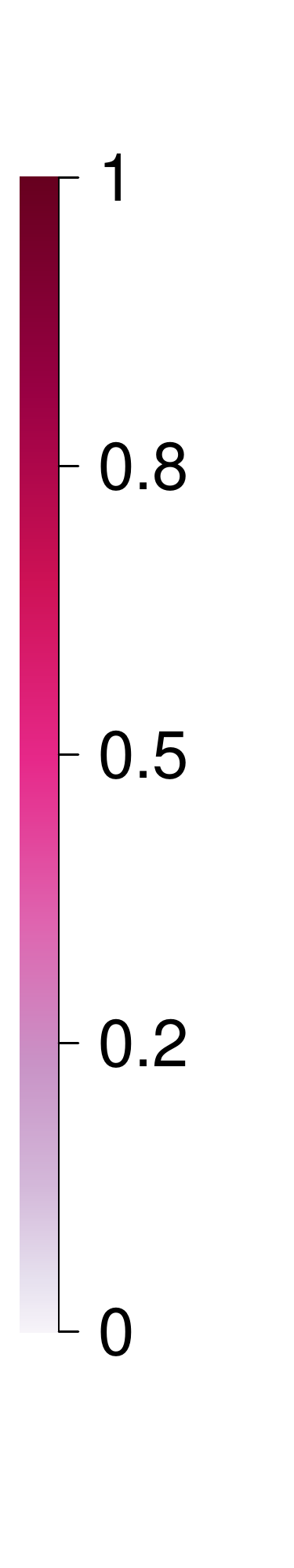}
}
	\caption{ITCZ location for January and July.}
	\label{fig:itcz}
\end{figure}

Let $\bm{\tilde{\mu}}_m=(\tilde{\mu}_{1m},\ldots,\tilde{\mu}_{lm},\ldots,\tilde{\mu}_{Lm})$ be the vector of the latent field predicted at locations $l=1,\ldots,L$, where $(lat_{1m},\ldots,lat_{lm},\ldots,lat_{Lm})$ is a regular sequence from $90^{\circ}$ to $-90^{\circ}$. 
%Formally, we sample from $\pi(\tilde{\mu}_{lm}|\bm y)$ for latitude values $l=1,\ldots,L$ and meridian $m$. 
%Our strategy underpin a particular definition of the ITCZ as the region of length $w$ where TCWV predicted field show highest values. 
The algorithm proceeds as follows. For $m=1,\ldots,M$:

\begin{itemize}
	\item evaluate the bases at locations $l=1,\ldots,L$, this gives a meridian-specific $L \times K$ dimensional basis matrix $\bm{\tilde{B}_m}$;
	\item sample $G$ realizations from the posterior predictive distribution (\ref{eq:pred}) by computing $\bm{\tilde \mu}^g_m=\alpha^g+\bm{\tilde{B}_m}  \hat{\bm{\beta}}^g$, $g=1,\ldots,G$; 
	\item for $g=1,\ldots,G$, rank the vector $\bm{\tilde \mu}^g_m$. This gives a posterior sample of the ranks, indicated by vector $\bm{\phi}_m^g=(\phi^{g}_{1m},\ldots,\phi^{g}_{lm},\ldots,\phi^{g}_{Lm})$, e.g. $\phi^{g}_{lm}=L$ if  $l=\argmax_{l}(\bm{\tilde\mu}^{g}_{m})$, while $\phi^{g}_{lm}=1$ if $l=\argmin_{l}(\bm{\tilde\mu}^{g}_{m})$.
%	\item 
\end{itemize}
The probability that a point $l$ belonging to meridian $m$ falls into the ITCZ is computed as 
\begin{equation}
Pr\left(lat_{lm} \in ITCZ|\bm y \right)=\frac{1}{G}\sum_{g=1}^{G} I\left(1-\frac{\phi^{g}_{lm}}{L} < w \right)
\label{eq:itcz-formula}
\end{equation}

where $I$ is the indicator function and $\phi^g_{lm}/L$ is the normalised rank. To sum up, (\ref{eq:itcz-formula}) is the probability that the point with geographical coordinates  $(lat_l,long_m)$ falls inside the ITCZ, where the length of the ITCZ is fixed according to $W$. %\textit{Setting $W=1000$, i.e. $w=0.05$, the ITCZ  (\ref{eq:itcz-formula}) expresses the probability that $(l,m)$ is in the $5\%$ area with highest values of the latent TCWV field.}
 Results are displayed in Figure \ref{fig:itcz} for the two months under study: this Figure is obtained running the algorithm with $L=1000$ and $M=360$. The ITCZ is mostly located in the south (north) of the Equator in January (July), as expected on the basis of prior knowledge concerning its seasonal behaviour. The map for January shows the double ITCZ, which is typical of the Central Pacific region in some period during the year \citep{waliser1993satellite}. The proposed method allows to locate the ITCZ even over land (in particular in Africa and South America) where data are not available, this being reflected by higher posterior uncertainty. 
Of course, the width of ITCZ reported in Figure \ref{fig:itcz} is strictly dependent on the choice of $W$: although this is very relevant when studying a single month, we believe that it is not such a crucial choice if the method is used for studying the spatio-temporal trend of the phenomenon. Indeed, in this case it would be important to keep $W$ fixed along the study period in order to ensure comparability among results.

%\begin{figure}
%\begin{minipage}{16cm}
%\includegraphics[width=0.5\textwidth]{fig/ITCZ_January.pdf}
%\includegraphics[width=0.5\textwidth]{fig/ITCZ_July.pdf}
%\end{minipage}
%\caption{ITCZ location for January and July.}
%\label{fig:itcz}
%\end{figure}

\section{Discussion}
\label{sec:discussion}

We presented a Bayesian hierarchical framework for smoothing data collected worldwide at a large number of locations. With respect to traditional methods, the proposed model accounts for geodesic distances between the data, thus overcoming the limitations of covariance functions for Euclidean spaces when applied to global datasets. 

The non-parametric model formulation proposed extends the Bayesian P-spline approach for smoothing worldwide collected data. Assuming the sphere as a representation of the Globe, the idea is to build a new basis of B-splines on a suitable geodesic grid while keeping the hierarchical model formulation of Bayesian P-splines, with the associated advantages in terms of flexibility and computation. Two key features of P-splines are maintained in the Geodesic P-spline model: (a) the use of local bell-shaped functions, e.g. the B-splines on the icomesh, that yield a sparse basis matrix; (b) the use of B-splines centred at equally-spaced knots, i.e. the nodes of the icomesh. Point (b) suggests that an IGMRF for regular locations is a sensible prior distribution for the spline coefficients, giving stable marginal variances as opposed to the standard P-spline model construction. 
%i.e. stationarity of the underlying spatial process over the spherical domain, a feature which cannot be guaranteed with standard P-spline model constructions.} 
Computational efficiency is due to (a) reduction of the latent field dimension, as the smoothing prior operates on the spline coefficients (low-rank smoother) and (b) fast MCMC based on sparse Cholesky factorization of the structure matrix of the full conditional for the latent field. This advantages allow for fast fitting of the model to data collected worldwide at a high-resolution. 

We applied the Geodesic P-spline model to TCWV data retrieved with the AIRWAVE algorithm at a huge number of locations on Earth.  The smoothing approach in this example is desirable as it allows estimation of the field at unmonitored locations. We provided inferential tools to locate the ITCZ based on ranking samples from the posterior distribution of the latent field, estimated at a fine grid over the Globe. Results are coherent with prior knowledge concerning ITCZ, indicating a shift towards southern regions in autumn and winter.

To apply the method the user is required to take decisions on mainly two critical aspects. The first regards the number of split-in-four iterations, $\nu$, which determines the total number of basis, $K$. The main point is to select a large enough $K$ in order to capture the spatial variability of the latent field. We believe this choice has to be made according to the application at hand and that more work is needed to investigate strategies valid in general. A practical rule of thumb would suggest to run a sensitivity analysis, increasing $\nu$ until computation of the basis gets impractical, then run multiple Geodesic P-splines defined on the GDGG associated to the selected $\nu$'s, finally compare these models according to some criterion suitable for the case study under examination. In our application we checked that $\nu=\{5,6\}$ essentially gave the same results regarding the ITCZ location, which is our goal, therefore we fixed $\nu=5$ for presentation of results. A second critical aspect, ubiquitous in any Bayesian analysis, is the choice of hyperpriors. Typically, the prior for the random walk precision, $\pi(\tau_{\beta})$, impacts more than the prior on the noise precision. We expect large impact of $\pi(\tau_{\beta})$ in situations where sample size is small compared to the number of parameters to estimate. In the case study on TCWV, the sample size available for estimating each spline coefficients is high enough, which makes the impact of $\pi(\tau_{\beta})$ very small. In the results presented in Section \ref{sec:results} we used a Gamma with shape $a=1$ and rate $b=5e-5$ for both $\tau_{\beta}$ and $\tau_{\epsilon}$, after checking that the posterior $\pi(\tau_{\beta}|\bm y)$ was unchanged under different choices of $a$ and $b$. 
 %(For the sake of comparison of the impact of the prior in the different models, the ICAR structure matrix was scaled, so that the prior on $\tau_{\beta}$ can be transferred through models with different icosphere resolutions.) 
We believe that controlling that the posterior learns from the data in the same way for different choice of the prior is a reasonable approach to test robustness of the Bayesian specification. On the topic of prior selection for variance parameters the literature is growing fast in the last decade; see, e.g., \cite{gelman-2006, pcprior} and reference therein.

The model can be extended in several directions, both on the methodological and applied side. In this paper we focused on an IGMRF structure for the spline coefficients equivalent to the ICAR used for lattice data, using the six surrounding knots as neighbours. Investigation of geodesic grids suitable for higher order IGMRF priors would be interesting. Another attractive research line is the extension towards a model based on nested B-splines, defined on a set of geodesic grids of different resolution, following \cite{nychka-2012}. In a fully Bayesian framework this requires careful hyperprior specification, as it is not clear how to prevent confounding between nested components. 

On the applied side, a future research line worth to be investigated is modelling the ITCZ based on different proxy variables, focusing the analysis on a wide temporal range, following the ideas in \citep{elisa-2017}. The application of Geodesic P-spline models to the 20 years of ATSR data will allow the investigation of ITCZ meridional migration trends. Moreover, joint modelling of TCWV and other ITCZ related phenomena, possibly available at misaligned locations, will result in more reliable estimates of the ITCZ latent field, especially at locations where TCWV retrieval is not possible with the current ATSR technology.

\section*{Acknowledgements}
The work by Fedele Greco and Massimo Ventrucci is funded by the PRIN2015 supported-project
\emph{Environmental processes and human activities: capturing their
interactions via statistical methods} (EPHASTAT) by MIUR
(Italian Ministry of Education, University and Scientific Research).
We thank Bianca Maria Dinelli (ISAC-CNR), Enzo Papandrea and Stefano Casadio (Serco s.p.a.) for guidance in the interpretation of the results. ATSR TCWV dataset was developed in the frame of the ESA ALTS project ESA Contract No. 4000108531/13/I-NB.

\bibliographystyle{apalike}
\bibliography{biblio}

\end{document}